\documentclass{aa}
\usepackage[varg]{txfonts}
\usepackage{graphicx}
\bibpunct{(}{)}{;}{a}{}{,} 

\begin{document}
\title{Jiamusi Pulsar Observations:  
  II. Scintillations of 10 Pulsars}

\author{P.~F. Wang\inst{1,2,5} \thanks{E-mail: pfwang@nao.cas.cn}
        \and J.~L. Han\inst{1,4,5} \thanks{E-mail: hjl@nao.cas.cn}
        \and Lei Han\inst{3}
        \and JianHui Zhang\inst{2,3}
        \and JunQiang Li\inst{2,3}
        \and \\ Chen Wang\inst{1,5}
        \and Jun Han\inst{1}
        \and Tao Wang\inst{1,4}
        \and X.~Y. Gao\inst{1,5}}
\institute{National Astronomical Observatories, Chinese Academy of Sciences,
           Jia-20 Datun Road, Chaoyang District, Beijing 100101, China
           \and
           The State Key Laboratory of Astronautic Dynamics, Xi'an, Shaanxi 710043, China
           \and
           Jiamusi Deep Space Station, China Xi'an Satellite Control Center,
           Jiamusi, Heilongjiang 154002, China
           \and
           School of Astronomy, University of Chinese Academy of Sciences,
           Beijing 100049, China
           \and
           CAS Key Laboratory of FAST, NAOC, Chinese Academy of Sciences, Beijing 100101, China
       }

\date{Received YYY / Accepted XXX}


\abstract
    %
    {Pulsars scintillate. Dynamic spectra show brightness variation 
      of pulsars in the time and frequency domain. Secondary spectra
      demonstrate the distribution of fluctuation power in the
      dynamic spectra.}
    %
    %
    {Dynamic spectra strongly depend on observational frequencies, but
      were often observed at frequencies lower than 1.5~GHz. Scintillation
      observations at higher frequencies help to constrain the
      turbulence feature of the interstellar medium over a wide
      frequency range and can detect the scintillations of more distant
      pulsars.}
    %
    %
    {Ten pulsars were observed at 2250~MHz (S-band) with the
    Jiamusi 66~m telescope to study their scintillations. Their dynamic
    spectra were first obtained, from which the decorrelation
    bandwidths and time scales of diffractive scintillation were then
    derived by autocorrelation. Secondary spectra were calculated
      by forming the Fourier power spectra of the dynamic spectra.}
    %
    %
    {Most of the newly obtained dynamic spectra are at the highest
    frequency or have the longest time span of any published data
      for these pulsars. For PSRs B0540+23, B2324+60 and B2351+61,
    these were the first dynamic spectra ever reported. The
    frequency-dependence of scintillation parameters indicates that
    the intervening medium can rarely be ideally turbulent with a
    Kolmogorov spectrum. The thin screen model worked well at S-band
    for the scintillation of PSR B1933+16. Parabolic arcs were
    detected in the secondary spectra of three pulsars, PSRs B0355+54,
    B0540+23 and B2154+40, all of which were asymmetrically
    distributed. The inverted arclets of PSR B0355+54 were seen
      to evolve along the main parabola within a continuous observing
    session of 12 hours, from which the angular velocity of the pulsar
    was estimated that was consistent with the measurement by very
    long baseline interferometry (VLBI).}
    %
     {}

\keywords{pulsars: general -- ISM: general -- pulsars: individual:
  PSRs B0329+54, B0355+54, B0540+23, B0740$-$28, B1508+55, B1933+16,
  B2154+40, B2310+42, B2324+60 and B2351+61}

\maketitle
\titlerunning{Scintillations of 10 pulsars at S-band}
\authorrunning{P. F. Wang, et al.}

\section{INTRODUCTION}

Pulsars are radio point sources and often move at high speeds of a few
tens to more than a thousand km~s$^{-1}$. When radio signals from
pulsars propagate through the interstellar medium, they are scattered
due to irregularly distributed thermal electrons. The scattering due
to the small scale irregularities of the medium can cause not only the
delayed arrival of the scattered radiation, shown as larger temporal
broadening of pulse profiles at lower frequencies, but also angular
broadening for the scattering disk of a pulsar image that can be
observed by VLBI. The random electron density fluctuations in the
interstellar medium can be quantitatively described by a power
spectrum of $P(k)=C_n^2 k^{-\beta}$ in a given spatial scale range of
$L$ for $k=2\pi/L$ in a given interstellar region \citep{ars95}; here,
$C_n^2$ is a measure of fluctuations. A Kolmogorov spectrum with
$\beta \simeq 11/3$ is widely used to describe the turbulent
medium. Nevertheless, the electron density fluctuations in the
interstellar medium, at least in some regions, do not follow the
Kolmogorov spectrum and may have a different spectral index $\beta$
\citep[e.g.][]{sg90,brg99c}.

\begin{table*}
  \caption{Parameters of 10 pulsars are listed together with frequencies for previous dynamic spectrum observations.}
  \label{tab1}
  \centering
  \begin{tabular}{lcrrrrllcc}
    \hline
    \hline
    PSRs & Period & DM &  \multicolumn{1}{c}{$l$}   &  \multicolumn{1}{c}{$b$} &  \multicolumn{1}{c}{$z$}  & Distance & $V_{\rm p}$  & \multicolumn{2}{c} {Freq. of Prev. Obs. (ref. Table~\ref{tab2})}   \\
     & (s) & (pc~cm$^{-3}$) & ($^{\circ}$) &($^{\circ}$) & (kpc) & (kpc) &$(\rm km~s^{-1})$ & $\nu$ < 2.3~(GHz) & $\nu \ge$ 2.3 (GHz) \\
    \hline
   B0329$+$54 & 0.714 & 26.76 & 145.00 & $-$1.22 & $-$0.02 & 1.0(1)$^d$ & 95(12)$^a$   & 0.327, 0.408, 0.61 & - \\
     &  &  &  &  &  & & & 0.96, 1.42, 1.54  &  \\
   B0355$+$54 & 0.156 & 57.14 & 148.19 &    0.81 &    0.01 & 1.0(2)$^d$ & 61(12)$^b$   & 0.325, 0.96 & - \\
   B0540$+$23 & 0.245 & 77.70 & 184.36 & $-$3.32 & $-$0.09 & 1.6(2$^\dagger$)$^e$ & 166(25$^\dagger$)$^f$ & -  & - \\
   B0740$-$28 & 0.166 & 73.78 & 243.77 & $-$2.44 & $-$0.09 & 2(1)$^d$   & 277(42$^\dagger$)$^f$ & -    & 4.8, 8.4 \\
   B1508$+$55 & 0.739 & 19.61 & 91.33  &   52.29 &    1.66 & 2.1(1)$^d$ & 963(64)$^c$  & 0.327,0.408 & - \\
   B1933$+$16 & 0.358 &158.52 & 52.44  & $-$2.09 & $-$0.14 & 3.7(13)$^d$& 394(208)$^c$ & 1.67 & - \\
   B2154$+$40 & 1.525 & 71.12 & 90.49  &$-$11.34 & $-$0.57 & 2.9(5)$^d$ & 264(73)$^c$  & 1.42  & - \\
   B2310$+$42 & 0.349 & 17.27 & 104.41 &$-$16.42 & $-$0.30 & 1.06(8)$^d$& 125(10)$^c$  & 0.327 & - \\
   B2324$+$60 & 0.233 &122.61 & 112.95 &    0.00 &    0.00 & 2.7(4$^\dagger$)$^e$ & 245(37$^\dagger$)$^f$ & - & - \\
   B2351$+$61 & 0.944 & 94.66 & 116.24 & $-$0.19 & $-$0.01 & 2.4(4$^\dagger$)$^e$ & 259(39$^\dagger$)$^f$ & - & - \\
   \hline                              
  \end{tabular}
  \tablefoot{
    \tablefoottext{a}{\citet{bbgt02}};
    \tablefoottext{b}{\citet{ccv+04}};
    \tablefoottext{c}{\citet{cbv+09}};
    \tablefoottext{d}{\citet{vwc+12}};
    \tablefoottext{e}{\citet{ymw17}};
    \tablefoottext{f}{\citet{mhth05}}.  
   $\dagger$ Uncertainties in brackets for the last digit are estimated to be 15\% of the values from http://www.atnf.csiro.au/research/pulsar/psrcat/. }
\end{table*}

A moving pulsar with a transverse velocity of $V_{\rm p}$ shines
through the relatively stable interstellar medium with small scale
irregularities of 10$^{6-8}$~cm causing diffractive
scintillation. This is exhibited by rapid fluctuations in time and
radio frequency in a dynamic spectrum. The typical time-scale, $\Delta
t_d$, and decorrelation bandwidth, $\Delta\nu_d$, depend on
observation frequency $\nu$ and the amount of intervening medium
indicated by $\rm DM$ or roughly by pulsar distance $D$, in the form
of \citep[see][]{ric77,wmj+05}
\begin{equation}
  \Delta t_d  \propto  \nu^{6/5} D^{-3/5} V_{\rm s}^{-1}, \\
  \Delta \nu_d \propto  \nu^{22/5} D^{-11/5}.
  \label{eq:parafreq}
\end{equation}
Here $V_{\rm s}$ is the speed of scintillation pattern past the
observer, which is caused by the velocities of the source and the
Earth as well as the intervening medium, and has often been used to
estimate the pulsar velocity $V_{\rm p}$ assuming other velocities are
negligible. The scintillation speed $V_{\rm s}$ can be estimated from
$\Delta t_d$, $\Delta \nu_d$ and pulsar distance $D$ by
\citep{ls82,cor86,gup95},
\begin{equation}
  V_{\rm s}= A \left(\frac{D}{\rm kpc}\right)^{1/2} \left(\frac{\Delta
    \nu_d}{\rm MHz}\right)^{1/2} \left(\frac{\nu}{\rm GHz}\right)^{-1}
  \left(\frac{\Delta t_d}{\rm s}\right)^{-1},
  \label{eq:viss}
\end{equation}
assuming a thin scattering screen. The constant $A$, depending on the
screen location and the form of the turbulence spectrum, was found to
be $A=3.85\times10^4~\rm km~s^{-1}$ \citep{gup95} for a screen located
at a half way between the pulsar and the observer.
The scintillation strength, $u$, originally defined as the ratio of
the Fresnel scale $s_F$ with respect to the field coherence scale
$s_0$ \citep[i.e, $u=s_F/s_0$, see][]{ric90}, can be estimated from
the observation frequency $\nu$ and the decorrelation bandwidth
$\Delta \nu_d$ by 
\begin{equation}
  u\simeq \left(\frac{\nu}{\Delta \nu_d}\right)^{0.5}.
  \label{eq:u}
\end{equation}
The fluctuation measure $C_n^2$ can be estimated from $\nu$, $\Delta
\nu_d$ and pulsar distance $D$ by \citep{cwb85}
\begin{equation}
  C_n^2   \approx  0.002 \; \nu^{11/3} D^{-11/6} \Delta \nu_d^{-5/6}
  \label{eq:cn2}
\end{equation}
for the Kolmogorov case. When the interstellar medium has a different
fluctuation power-law spectrum, i.e. different index $\beta$, these
above scaling relations could be different \citep{rnb86,ric90,brg99c}.

When pulsar signals propagate through large-scale irregular clouds of
10$^{10-12}$~cm with various electron density distributions,
refractive scintillation can be observed as pulsar flux density
fluctuations on longer time scales in addition to small-amplitude
intensity variations \citep{rcb84}. Refraction moves scintillation
pattern laterally and causes its systematic drift, which is manifested
as fringes on pulsar dynamic spectra. The slope of the fringes is
\citep[e.g.][]{sw85,brg99c}
\begin{equation}
  \frac{{\rm d} t}{{\rm d} \nu} \simeq \frac{D \theta_{r}}{V_{\rm s}} \frac{1}{\nu},
\end{equation}
where $\theta_r$ is the refractive angle which is approximately
proportional to $\nu^{-2}$ for a given gradient of refractive index.

Previously, scintillations of more than 80 pulsars have been observed
mostly at lower frequencies \citep[e.g.][]{ls82, ra82,sw85, bk85,
  cwb85,cor86, cw86,grl94,mss+96, brg99a, gg00,wmj+05}. For a given
pulsar, the decorrelation bandwidth $\Delta \nu_d$ and scintillation
time-scale $\Delta t_d$ are closely related to observation
frequenciess \citep[e.g.][]{cwb85}. Such observed frequency
dependencies can be used to estimate the power-law index $\beta$ for
electron density fluctuations in the interstellar medium, which has
often been found to deviate from the Kolmogorov spectrum \citep{grl94,
  brg99c, wmj+05}.

\begin{table*}
  \centering
  \caption{Previous observations of the 10 pulsars for scintillation parameters. }
  \label{tab2}
  \small
  \begin{tabular}{lrllllrlll}
    \hline
    \hline
    PSRs & Freq. &$\Delta \nu_d$&$\Delta t_d$& Reference  &    PSRs & Freq. &$\Delta \nu_d$&$\Delta t_d$& Reference    \\
         & (MHz) & (MHz)  & (min)  &                 &         & (MHz) & (MHz)  & (min)  &            \\
        \hline
B0329$+$54& 327  & 0.165  & 5.12    & \citet{brg99a} &  B0540$+$23 & 430 & 0.0019 &  -     & \citet{cwb85}     \\
         &  327  &0.02$^\dagger$ & - & \citet{wol83} &             & 1380 & 0.150  &  -     & \citet{cwb85}     \\
         &  330  &0.016$^\dagger$& - & \citet{wbs81} &             & 1410 & 0.31   &  -     & \citet{cwb85}     \\
         &  340  &0.023$^\dagger$& - & \citet{ar81}  &             & 1420 & 0.28   &  -     & \citet{cwb85}     \\
         &  408  & 0.047  &  3.23  & \citet{grl94}   &             & 1420 & 0.317  &  -     & \citet{cwb85}   \\
         &  408  & 0.083  &  4.5   & \citet{ls82}    &             & 4750 &  -     &  8     & \citet{mss+96}  \\
         &  410  &0.056$^\dagger$& - & \citet{ar81}  &             &10550 &  -     &  8     & \citet{mss+96}    \\
         &  410  &0.07$^\dagger$ & - & \citet{ric70} &  B0740$-$28 &  660 &  -    &  0.97   & \citet{jnk98}    \\
         &  410  & 0.100  &  -     & \citet{ric77}   &             & 4750 &  -     &  5      & \citet{mss+96} \\
         &  480  & 0.103  &  -     & \citet{wol83}   &             & 4800 &  8.83  & 10.63   & \citet{jnk98}  \\
         &  610  & 0.13   &  4.43  & \citet{spg+17}  &             & 8400 & 40.0   & 37.67   & \citet{jnk98}  \\
         &  610  & 0.22   &  5.34  & \citet{spg+17}  &             & 10550 &  -    & 3.5     & \citet{mss+96} \\
         &  610  &0.348$^\dagger$& - & \citet{ric70} &  B1508$+$55 &  327 & 0.168  & 2.63   & \citet{brg99a}   \\
         &  610  & 0.349  &  5.90  & \citet{sfm96}   &             &  327 & 0.226  & 2.73    & \citet{brg99a} \\
         &  960  & 0.92   & 12.94  & \citet{sw85}    &             &  340 & 0.139$^\dagger$& - & \citet{ar81}  \\
         & 1410  & 2      &  -     & \citet{whs74}   &             &  408 &  0.8    &  3     & \citet{sw85}  \\
         & 1420  & 5.93   & 15.52  & \citet{spg+17}  &             &  408 &  1.67   &  -     & \citet{ls82}  \\
         & 1540  & 14     & 16.9   & \citet{wmj+05}  &             &  410 & 0.13$^\dagger$ & - & \citet{ric70} \\
         & 1540  & 9.2    & 17.1   & \citet{wymw08}  &             &  930 &  -      & 1.35   & \citet{ls82}  \\
         & 4750  &  -     & 21     & \citet{mss+96}  &  B1933$+$16 & 1410 &  0.125 &  -     & \citet{whs74}  \\
         & 4800  &  -     & 42.7   & \citet{lkgk11}  &             & 1416 & 0.037$^\dagger$& - & \citet{wol83} \\
         &10550  &  -     & 23     & \citet{mss+96}  &             & 1420 & 0.1    &  -     & \citet{ric77}  \\
B0355$+$54& 325  & 0.06   &  1.01  & \citet{spg+17}  &             & 1670 & 0.110  & 0.75   & \citet{ra82}   \\
          &  408  & -      &  1.83  & \citet{ls82}   &  B2154$+$40 & 1000 &  0.195 &  0.503 & \citet{cor86}  \\
          &  410  &0.018$^\dagger$& - & \citet{ar81} &             & 1420 &  0.20  &  0.55 & \citet{spg+17}   \\
          &  930  & 0.765  &  -     & \citet{ls82}   &  B2310$+$42 &  327 &  0.114 &  5.15  & \citet{brg99a}\\
          &  960  & 0.613  &  4.31  & \citet{sw85}   &  B2324$+$60 &  -   &  -     &  -     &    -           \\
          & 1410  & 0.575  &  -     & \citet{whs74}  &  B2351$+$61 &10550 &  -     & 15     & \citet{mss+96} \\
          & 4750  &  -     & 12.5   & \citet{mss+96} &             &      &        &        &     \\ 
          & 10550 &  -     & 13.5   & \citet{mss+96} &             &      &        &        &    \\
\hline
  \end{tabular}
  \tablefoot{$\dagger$ decorrelation bandwidths rescaled to
    the half-maximum of the correlation functions according to
    \citet{cwb85} by assuming Kolmogorov irregularity for the
    interstellar medium. }
\end{table*}

When a high-sensitivity dynamic spectrum of a pulsar is obtained, not
only strong large patterns in time and bandwidth are observed; but
also faint organized structures on smaller scales may appear in the
dynamic spectrum image. These are best studied through the secondary
spectrum, which is the power spectrum of the dynamic spectrum;
i.e. $S_2 (f_t, f_\nu) = |\tilde{S_1} (t, \nu)|^2$, where $S_1$ is the
dynamic spectrum and the tilde indicates a Fourier transform. Assuming
that the pulsar velocity dominates and that any linear scattering
structure is aligned along the effective velocity vector, one can
estimate the fractional distance $d$ of the intervening screen from a
pulsar at the distance $D$ with a velocity $V_{\rm p}$ based on the
curvature of the arc in the secondary spectrum via \citep[see
  e.g.][]{crsc06}
\begin{equation}
   f_{\nu} = \frac{\lambda^2 D}{2cV_{\rm p}^2} \left(\frac{d}{1-d}\right) f_{t}^2 ,
   \label{eq:secspec}
\end{equation}
here, $f_{t}=1/t$ is the conjugate time, $f_{\nu}=1/\nu$ is conjugate
frequency, $\lambda$ is the observing wavelength, $c$ is the speed of
light. The curvature of a primary arc in the $f_{t}$ and $f_{\nu}$
plane is
\begin{equation}
  \Omega = \frac{\lambda^2 D}{2cV_{\rm p}^2} \left(\frac{d}{1-d}\right).
\end{equation}
Previously, such arcs in the secondary spectra have been detected for
only 13 pulsars: PSRs B1133+16 \citep{cw86,smc+01,hsb+03}, B0823+26,
B0834+06 \citep{smc+01,hsb+03}, B0919+06 and B1929+10
\citep{smc+01,hsb+03, sti07}, J0737$-$3039 \citep{shr05}, B1737+13
\citep{crsc06,sti07}, B0355+54 \citep{sti07, xlh+18}, J0437$-$4715
\citep{bot+16}, B1642$-$03, B1556$-$44, B2021+51 and B2154+40
\citep{spg+17}. The placements of the intervening screen have so been
estimated.

In this paper, we present the scintillation observations of ten
pulsars using the Jiamusi 66-m telescope at 2250~MHz. Parameters of
these pulsars are listed in Table~\ref{tab1}, and previous
scintillation observations are given in Table~\ref{tab2}. The dynamic
spectra presented in this paper are valuable supplements to the
previous observations and, in some cases, are the first dynamic and
secondary spectra ever published. In Section 2 we describe our
observation system. Observational results are presented and analyzed
in Section 3. Discussion and conclusions are given in Sections 4 and
5, respectively.

\begin{table*}
  \centering
  \caption{Observational parameters and derived scintillation
    parameters, with uncertainties for the last digit given in
    brackets.}
  \label{tab3}
  \tabcolsep 1.5mm
  \begin{tabular}{llcc crrr rrlr}
    \hline
    \hline
    PSR name & \multicolumn{1}{c}{Date} & MJD &chW &$\rm \delta t$&$T_{\rm obs}$ &\multicolumn{1}{c}{$\Delta \nu_d$} &
    \multicolumn{1}{c}{$\Delta t_d$} & 
    ${\rm d}t/{\rm d}\nu$ & \multicolumn{1}{c}{$u$} &   log $C_n^2$  & \multicolumn{1}{c}{$ V_{\rm s}$} \\
             & &      & (MHz)  & (s)  & (min) & (MHz) & (min) & (min/MHz) &  &$(\rm m^{-20/3})$& $(\rm km~s^{-1})$\\
    \hline
   B0329$+$54 & 2016/02/21 & 57439.103 & 0.58 & 30 & 306 &  17(2)  & 19(2)    & -0.09(1)  & 12(1) & -2.43(10) &  62(7)  \\
              & 2016/02/24 & 57442.227 & 0.58 & 30 & 426 &  20(2)  & 26(3)    & -0.39(5)  & 11(1) & -2.49(8) &  49(6) \\ 
              & 2017/11/08 & 58065.445 & 0.58 & 30 & 655 &  67(14) & 30(6)    &  0.02(1)  &  6(1) & -2.93(17)&  78(18) \\ 
   B0355$+$54 & 2015/08/20 & 57254.546 & 0.58 & 30 & 654 &   15(1) &  9(1)    &  0.18(1)  & 12(1) & -2.39(4) & 123(14) \\
              & 2016/01/29a& 57416.036 & 0.58 & 30 & 720 &  4.0(1) & 2.4(1)   & -0.39(1)  & 24(1) & -1.91(2) & 238(10) \\ 
              & 2016/01/29b& 57416.556 & 0.58 & 30 &  96 &  4.3(2) & 2.9(1)   & -0.35(2)  & 23(1) & -1.94(4) & 204(8) \\
              & 2017/11/05 & 58062.678 & 0.58 & 30 & 486 &  41(5)  & 13(2)    & -0.23(2)  &  8(1) & -2.75(10)& 140(23)\\
              & 2017/11/09 & 58066.421 & 0.58 & 30 & 362 &  22(2)  &  7(1)    & -0.16(1)  & 10(1) & -2.53(8) & 191(29)\\
   B0540$+$23 & 2015/06/25 & 57198.154 & 0.58 & 30 &  50 &  1.9(1) & 2.9(1)   &  0.95(5)  & 35(1) & -1.99(4) & 169(7)\\ 
              & 2016/08/08 & 57608.863 & 0.58 & 30 & 168 &  2.3(1) & 2.5(1)   &  0.53(4)  & 32(1) & -2.06(4) & 216(10) \\ 
              & 2017/11/01 & 58058.549 & 0.58 & 30 & 336 &  1.3(1) & 2.6(1)   &  0.30(2)  & 42(2) & -1.86(6) & 156(8) \\ 
   B0740$-$28 & 2015/12/12 & 57368.687 & 0.58 & 30 & 168 &  1.1(1) & 5.2(2)   & -0.05(12) & 45(2) & -1.99(8) &  81(5) \\ 
              & 2016/01/26 & 57413.544 & 1.17 & 30 & 186 &  1.1(1) & 2.9(1)   &  0.20(14) & 45(2) & -1.99(8) & 146(8) \\ 
              & 2017/11/01 & 58058.793 & 0.58 & 30 & 156 &  0.6(1) & 3.0(1)   & -0.54(19) & 61(5) & -1.77(14) & 104(9) \\ 
   B1508$+$55 & 2015/12/13 & 57369.830 & 0.58 & 60 & 138 & 8.1(5)  & 3.2(2)   & -0.20(1)  & 17(1) & -2.76(5) & 368(26)\\
              & 2017/10/31 & 58057.067 & 0.58 & 60 &  67 & 44(8)   &   4(1)   & -0.03(1)  &  8(1) & -3.37(15)& 685(182)\\
   B1933$+$16 & 2015/06/15 & 57188.756 & 0.58 & 30 &  31 & 1.36(6) & 2.35(9)  &  0.74(5)  & 41(1) & -2.56(4) & 272(12)\\
              & 2016/02/18 & 57436.861 & 0.58 & 30 & 300 & 0.97(2) & 1.60(5)  &  0.37(7)  & 48(1) & -2.44(2) & 338(11)\\ 
              & 2016/02/25 & 57443.861 & 0.58 & 30 & 564 & 0.77(3) & 1.37(6)  &  0.24(11) & 54(1) & -2.35(1) & 351(17)\\ 
              & 2016/05/20 & 57528.742 & 0.58 & 30 & 378 & 1.82(2) & 2.03(2)  & -0.26(2)  & 35(1) & -2.67(1) & 365(4)\\ 
              & 2016/05/21 & 57529.629 & 0.58 & 30 & 540 & 1.67(1) & 2.01(2)  & -0.24(2)  & 37(1) & -2.63(1) & 353(4)\\ 
              & 2016/05/22 & 57530.652 & 0.58 & 30 & 504 & 1.51(2) & 2.03(2)  & -0.39(2)  & 39(1) & -2.60(1) & 332(4)\\ 
              & 2016/05/23 & 57531.730 & 0.58 & 30 & 389 & 1.79(2) & 2.20(3)  & -0.25(2)  & 36(1) & -2.66(1) & 334(5)\\ 
              & 2016/05/25 & 57533.790 & 0.58 & 30 & 200 & 2.58(6) & 2.61(6)  & -0.16(2)  & 30(1) & -2.79(2) & 338(9)\\ 
   B2154$+$40 & 2016/01/25 & 57412.291 & 0.58 & 30 & 164 & 1.60(3) & 1.42(3)  & -0.26(2)  & 38(1) & -2.43(2) & 433(10)\\
              & 2017/10/31 & 58057.533 & 0.58 & 30 & 276 & 2.34(5) & 2.44(5)  & -0.68(3)  & 31(1) & -2.56(2) & 304(7)\\ 
              & 2017/11/10 & 58067.151 & 0.58 & 30 & 216 & 4.1(1)  & 2.41(7)  & -0.27(2)  & 24(1) & -2.77(2) & 408(13)\\
   B2310$+$42 & 2015/07/20 & 57223.566 & 0.58 & 90 & 145 & 15(2)   & 13(2)    & -0.51(7)  & 13(1) & -2.43(11)&  87(15)\\
              & 2017/10/25 & 58051.568 & 0.58 & 30 & 360 & 35(6)   & 23(4)    & -0.27(5)  &  8(1) & -2.74(14)&  76(15)\\
   B2324$+$60 & 2017/03/29 & 57841.268 & 0.58 & 60 & 187 & 2.3(1)  & 1.7(1)   &  0.13(2)  & 32(1) & -2.51(4) & 420(26)\\
              & 2017/11/04 & 58061.117 & 0.58 & 60 & 346 & 2.2(1)  & 1.7(1)   &  0.33(2)  & 32(1) & -2.49(4) & 411(26)\\ 
   B2351$+$61 & 2015/06/18 & 57191.677 & 0.58 & 90 &  35 & 2.5(2)  & 2.6(2)   & -0.38(2)  & 30(1) & -2.44(7) & 269(23)\\
              & 2016/08/10 & 57610.886 & 0.58 & 90 & 150 & 6.0(3)  & 2.9(2)   &  0.15(1)  & 20(1) & -2.75(4) & 373(27)\\
   \hline                                                                                           
  \end{tabular}
\end{table*}

\section{Observations and data processing}

Observations of ten pulsars were carried out between 2015 June and
2017 November using the Jiamusi 66-m telescope at the Jiamusi Deep
Space Station, China Xi'an Satellite Control Center. The observation
system used here is the same as the one described in \citet[][see
  their figure 1 for the diagram]{hhp+16}. In short, the Jiamusi 66-m
telescope is equipped with a cryogenically cooled dual-channel S-band
receiver. We observed pulsars with this receiver at the central
frequency of 2250~MHz with a bandwidth of about 140~MHz. The
down-converted intermediate frequency signals from the receiver for
the left and right hand polarizations were fed into a digital
backend. The signals were sampled and then channelized by an FFT
module in the digital backend. The total radio power from the two
polarizations was then added for each of the 256 frequency channels
with a channel width of 0.58~MHz. The data were saved to disk with a
time resolution of 0.2~ms. Alternatively, 128 channels were sampled
with a time resolution of 0.1~ms and with a channel width of
1.17~MHz. Observational parameters are listed in Table~\ref{tab3} for
the ten pulsars.

Offline data processing includes several steps. First, radio frequency
interference was manually identified from the two-dimensional plots of
data on the frequency channel and time domain, and the affected data
were simply excised. Second, the sampled total power data from every
channel were re-scaled according to the observations of the
flux-calibrators, 3C286 or 3C295. Third, data from each channel were
then folded with the ephemerides of the pulsars, with a subintegration
time of $\delta t =$ 30~s or 60~s or 90~s for a significant detection
of pulse flux density. For each subintegration at each frequency
channel, the flux density within the pulse window was calculated after
subtracting a baseline offset and then normalized by the offset to
form the dynamic spectrum, $S_1(t,\nu)$. During the analysis, software
packages: DSPSR \citep{vb11} and PSRCHIVE \citep{hvm04}, were
employed.

To obtain the auto-correlation function, the mean flux density
$\langle S_1 \rangle$ was first subtracted from the dynamic spectrum
of each observation to get $\Delta S_1(t,\nu)=S_1(t,\nu)- \langle S_1
\rangle$. Meanwhile, the previously excised pixels of the dynamic
spectrum were interpolated by nearby samples. Then, the covariance
function was computed \citep{cor86,wmj+05}:
\begin{equation}
  \begin{split}
   & C_f(\Delta t,\Delta \nu) = N(\Delta t,\Delta \nu)^{-1} \times
    \\ & \sum_{j=1}^{N_\nu-N_{|\Delta \nu|}}
    \sum_{i=1}^{N_t-N_{|\Delta t|}} \Delta S(t_i,\nu_j) \Delta
    S(t_i+\Delta t,\nu_j+\Delta \nu) .
   \end{split}
\end{equation}
Here, $N_\nu$ and $N_t$ are the numbers of frequency channels and
subintegrations respectively, $N_{|\Delta \nu|}$ and $N_{|\Delta t|}$
are the lags in the $\nu$ and $t$ directions, and $N(\Delta t,\Delta
\nu)$ represents the number of pairs for the correlation
pixels. Finally, the covariance function was normalized by its
amplitude at zero lags to get the autocorrelation function
\begin{equation}
  A(\Delta t,\Delta \nu)=\frac{C_f(\Delta t,\Delta \nu)}{C_f(0,0)}.
\end{equation}
Since $C_f(0,0)$ contains a significant ``noise'' from interpolated
dummy pixels, it was specially replaced by the peak value of the
Gaussian function fitted to nearby points (see below).

The scintillation parameters were derived from the auto-correlation
functions. Following \citet{grl94}, a two-dimensional elliptical
Gaussian function in the form of
\begin{equation}
  G(t,\nu)=C_0 {\rm exp}[-(C_1 \nu^2+C_2 \nu t+C_3 t^2)]
\end{equation}
was fitted to the main peak of $A(\Delta t,\Delta \nu)$ with $C_0$
fixed to unity. The decorrelation frequency, $\Delta \nu_d$, was
measured from the fitted two-dimensional elliptical Gaussian function
as the half width at the half maximum of $A(\Delta t,\Delta \nu)$, and
the time scale, $\Delta t_d$, was measured as the half width for
$A(\Delta t,\Delta \nu)$ to declined to $1/e$ \citep[see][]{cor86};
hence,
$ 
  \Delta \nu_d =(\ln2/C_1)^{0.5},
$ 
$ 
  \Delta t_d = (1/C_3)^{0.5},
$ 
and the orientation of the elliptical Gaussian was given by
$ 
   {{\rm d}t}/{{\rm d}\nu} =-C_2/(2C_3),
$ 
see \citet{bgr99b}. The so-deduced scintillation parameters for
observations of the ten pulsars are listed in
Table~\ref{tab3}. Meanwhile, as noted in \citet{bgr99b}, the
uncertainties in $\Delta \nu_d$, $\Delta t_d$ and ${{\rm d}t}/{{\rm
    d}\nu}$ consist of not only the fitting uncertainty but also the
statistical uncertainty caused by the finite number of independent
scintles in the dynamic spectra. The fractional uncertainty of scintle
statistics is given by
 \begin{equation}
   \sigma_{\rm est}=\left[f\left(\frac{B_{\rm obs}}{\Delta \nu_d}
     \frac{T_{\rm obs}}{\Delta t_d}\right)\right]^{-0.5},
 \end{equation}
here $B_{\rm obs}$ and $T_{\rm obs}$ represent the total observing
bandwith and time, and the effective filling factor $f$ for the
scintles is chosen to be $f=0.5$. In this paper, the fitting and
statistical uncertainties were added in quadrature to derive the
uncertainties of $\Delta \nu_d$, $\Delta t_d$ and ${{\rm d}t}/{{\rm
    d}\nu}$, and also derived parameters $u$, $C_n^2$ and $V_{\rm s}$,
in Table~\ref{tab3}, given in parentheses after each value.

The secondary spectrum was obtained by calculating the power spectrum
of the dynamic spectrum through two-dimensional FFT. During our
calculation, the dynamic spectra were first interpolated over the RFI
affected pixels to reduce the leakage of power. The generalized Hough
transform \citep[see details in][]{bot+16} together with fitting
parabolas by eye were used to obtain the curvatures of the secondary
spectra.

\begin{figure*}
  \begin{tabular}{l}
    \includegraphics[bb = 25 40 282 182, clip, height=0.15\textheight] {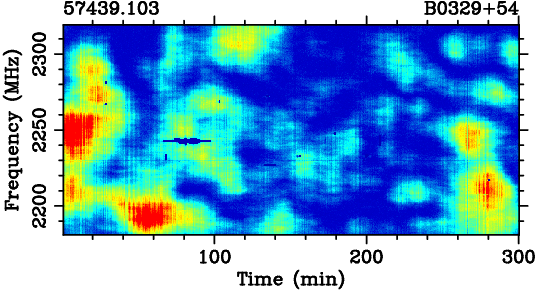} 
    \includegraphics[bb = 50 40 489 182, clip, height=0.15\textheight] {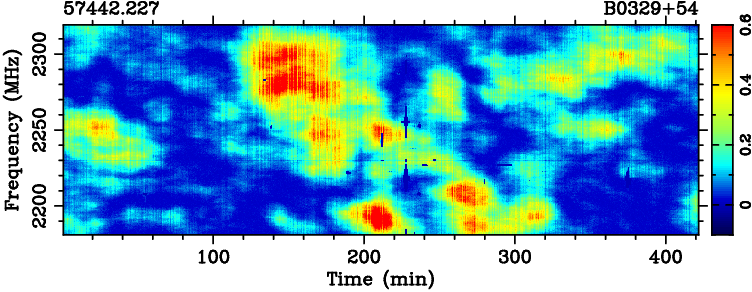} \\
    \includegraphics[bb = 25 40 530 182, clip, height=0.15\textheight] {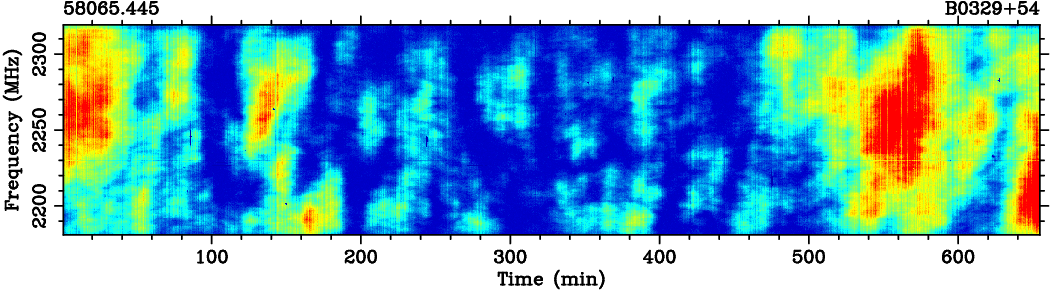} \\
    \includegraphics[bb = 32 32 280 272, clip, height=0.15\textheight] {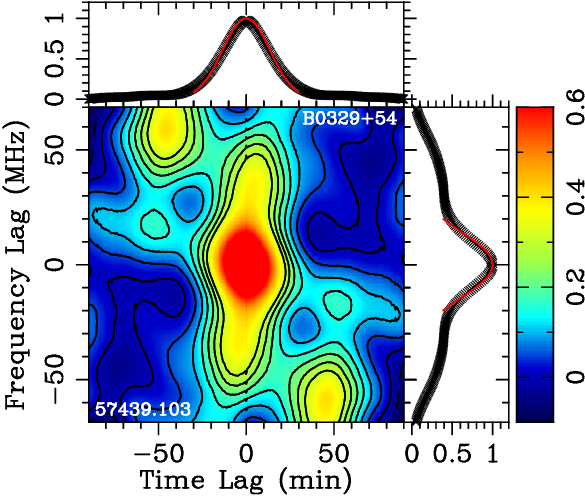} 
    \includegraphics[bb = 65 32 280 272, clip, height=0.15\textheight] {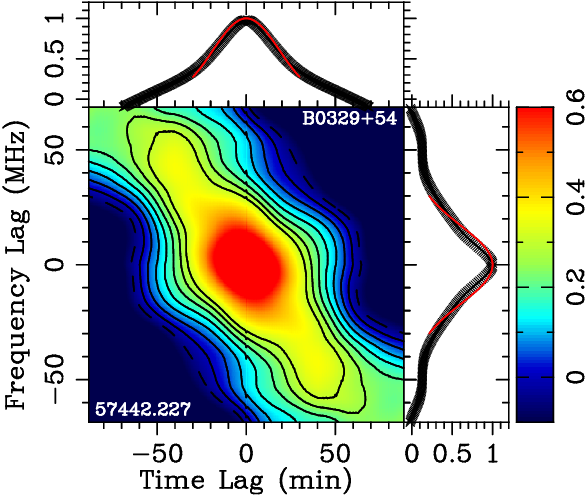}
    \includegraphics[bb = 65 32 320 272, clip, height=0.15\textheight] {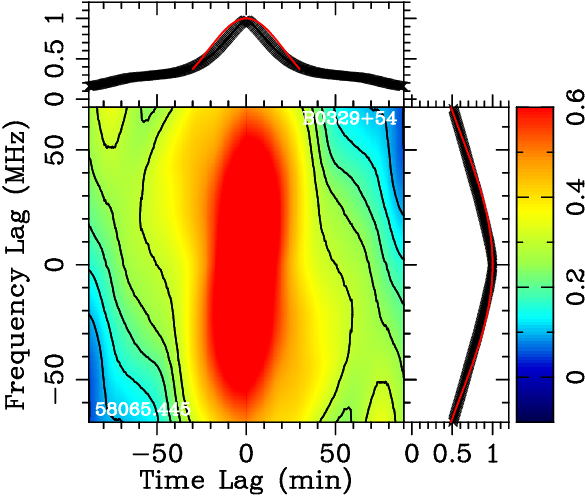}
   \end{tabular}
  \caption{The dynamic spectra $S_1(t,\nu)$ from three observations of
    PSR B0329+54 ({\it upper three panels}) and their auto-correlation
    functions $A(\Delta t,\Delta \nu)$ ({\it panels in the lowest
      rank}). Some dummy points caused by the interference are set as
    $S_1(t,\nu)=0$. Observation time (in MJD) and pulsar name are
    indicated above each plot. The auto-correlation functions are
    plotted against time lag $\Delta t$ in minutes and frequency lag
    $\Delta \nu$ in MHz. The two-dimensional Gaussian functions are
    fitted to the main peaks of auto-correlation functions to derive
    the scintillation parameters. In the side-panels only data for the
    cross-sections are shown by ``x'' (too many for this pulsar as if
    a thick line, but not so many for other pulsars shown in other
    figures) and the fitted lines at $\Delta t =0$ and $\Delta \nu =0$
    are drawn. The color schemes are indicated by the wedges for both
    the dynamic spectra and auto-correlation functions, which are
    saturated at $-10$\% and 60\% of the peak values (with exceptions
    for other pulsars but especially marked).}
  \label{fig0329}
\end{figure*}
\begin{figure*}
  \begin{tabular}{l}
    \includegraphics[bb = 25 40 510 182, clip, angle=0, height=0.15\textheight] {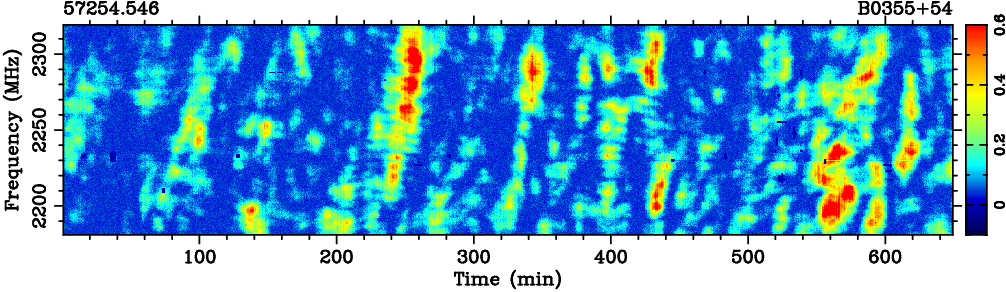}
    \includegraphics[bb = 50 40 166 182, clip, angle=0, height=0.15\textheight] {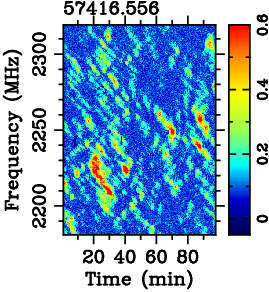}\\
    \includegraphics[bb = 25 40 529 182, clip, angle=0, height=0.15\textheight] {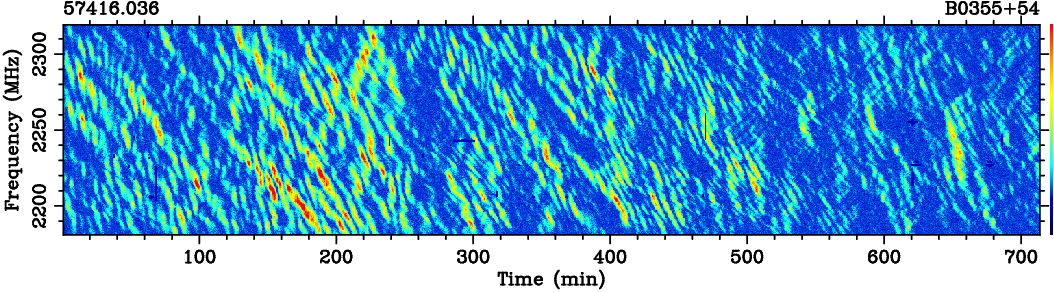}
    \includegraphics[bb = 488 40 510 182, clip, angle=0, height=0.15\textheight] {B0355+54_201508202106_ds.ps}\\
    \includegraphics[bb = 25 40 400 182, clip, angle=0, height=0.15\textheight] {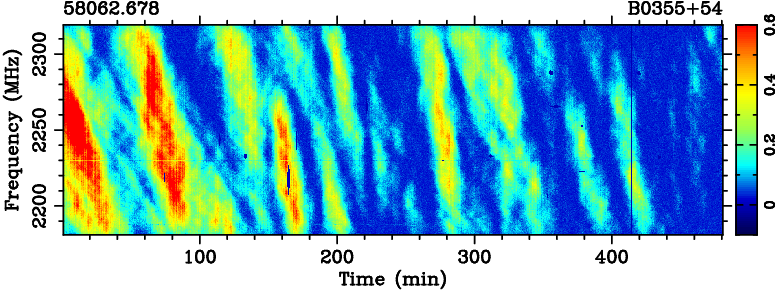}
    \includegraphics[bb = 50 40 400 182, clip, angle=0, height=0.15\textheight] {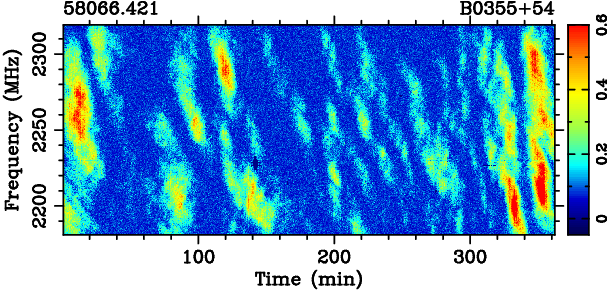}\\
    \includegraphics[bb = 32 32 279 272, clip,angle=0, height=0.15\textheight] {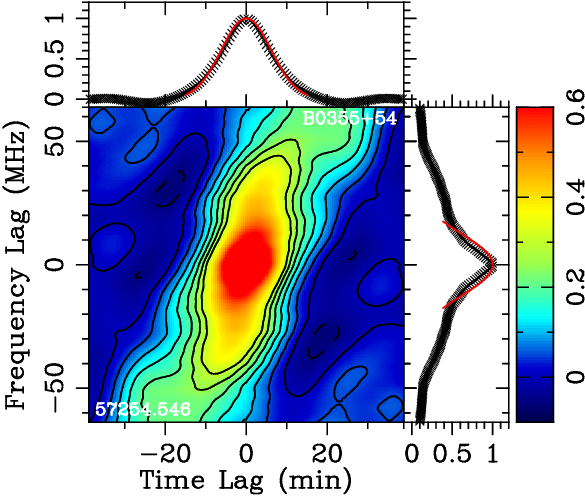}
    \includegraphics[bb = 65 32 279 272, clip,angle=0, height=0.15\textheight] {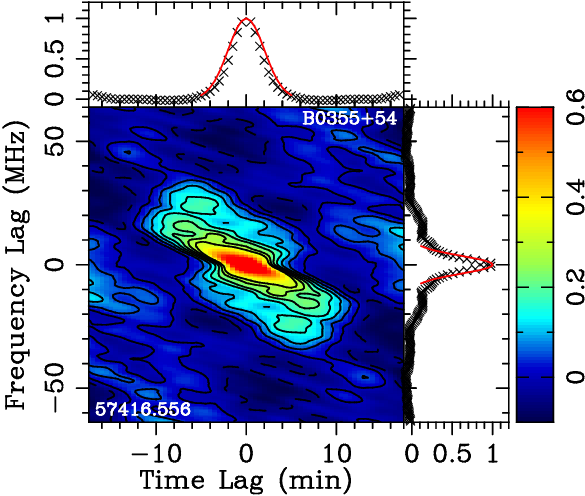} 
    \includegraphics[bb = 65 32 279 272, clip,angle=0, height=0.15\textheight] {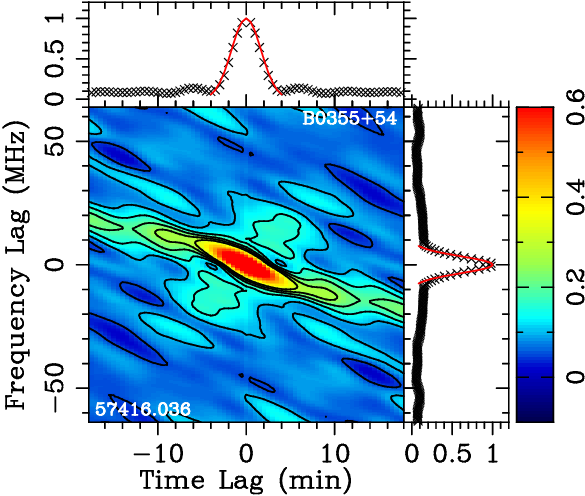}
    \includegraphics[bb = 65 32 279 272, clip,angle=0, height=0.15\textheight] {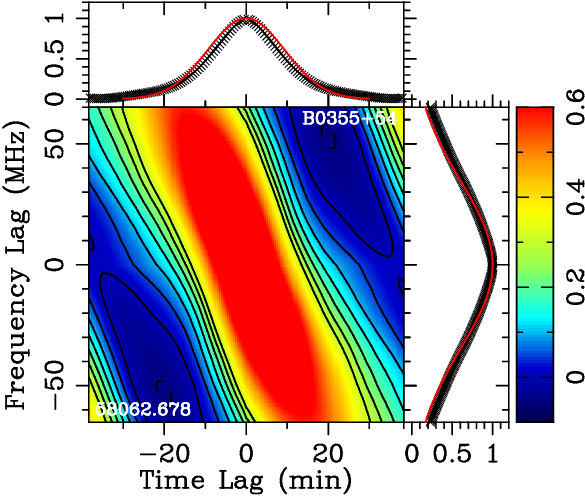}
    \includegraphics[bb = 65 32 320 272, clip,angle=0, height=0.15\textheight] {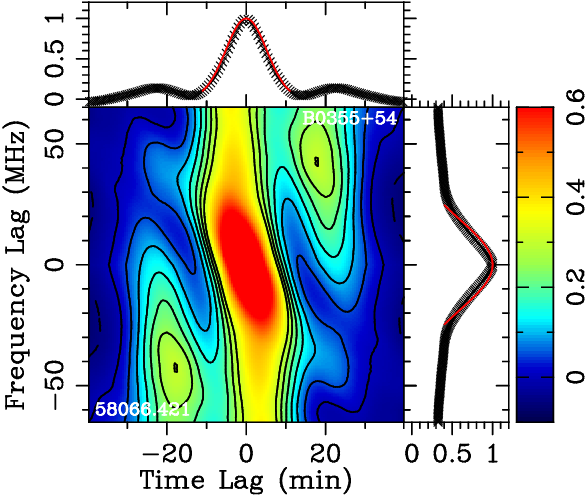}\\ 
  \end{tabular}
  \caption{Dynamic spectra from five observations of PSR B0355+54
    ({\it upper five panels}) and their autocorrelation functions
    ({\it panels in the last rank}). See keys in Fig.~\ref{fig0329}.
  }
  \label{fig0355}
\end{figure*}
\begin{figure*}
    \includegraphics[bb = 25 40 119 182, clip, angle=0, height=0.155\textheight] {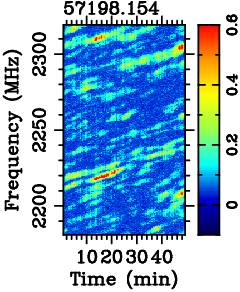}
    \includegraphics[bb = 50 40 231 182, clip, angle=0, height=0.155\textheight] {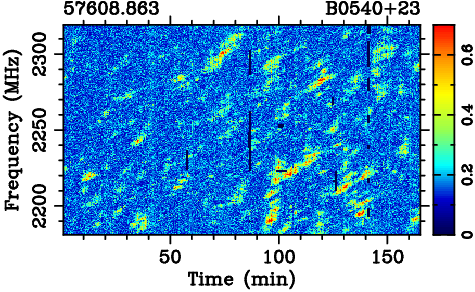} 
    \includegraphics[bb = 50 40 430 182, clip, angle=0, height=0.155\textheight] {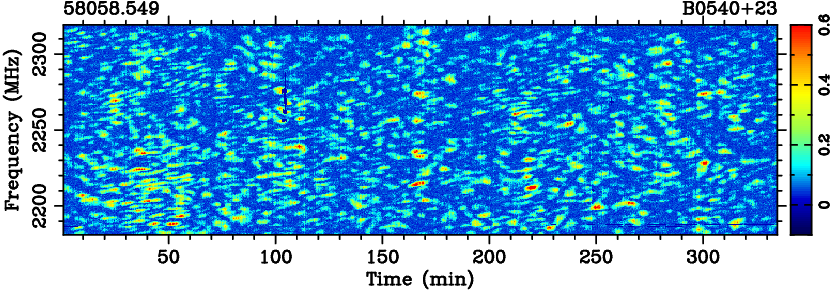} \\
    \includegraphics[bb = 32 32 279 272, clip, angle=0, height=0.154\textheight] {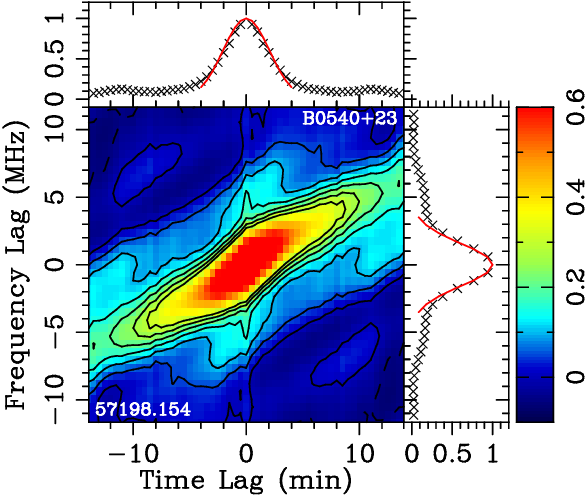}
    \includegraphics[bb = 65 32 279 272, clip, angle=0, height=0.154\textheight] {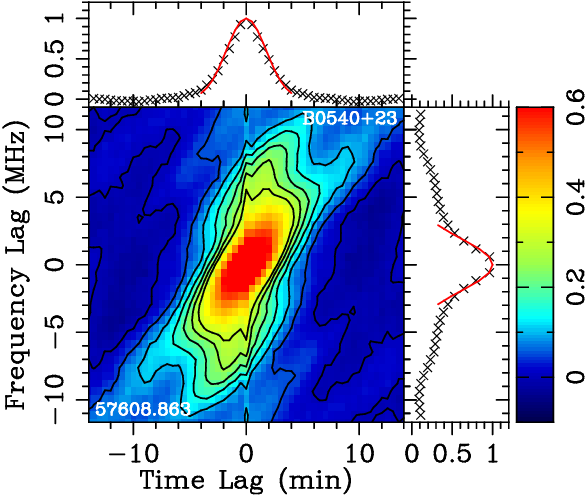}
    \includegraphics[bb = 65 32 320 272, clip, angle=0, height=0.154\textheight] {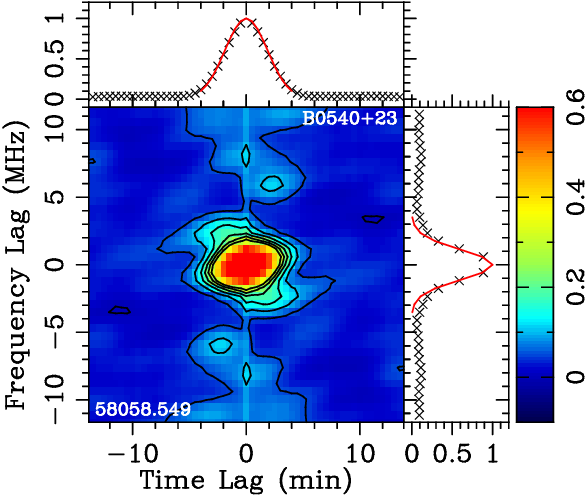}
  \caption{Dynamic spectra from three observations of PSR B0540+23
    ({\it upper panels}) and their autocorrelation functions ({\it
      lower panels}) See keys in Fig.~\ref{fig0329}.}
  \label{fig0540}  
\end{figure*}

\begin{figure*}
    \includegraphics[bb = 25 40 243 182, clip, angle=0, height=0.155\textheight] {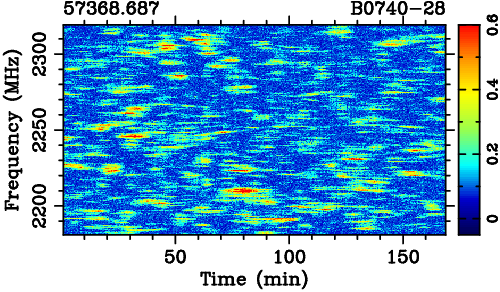} 
    \includegraphics[bb = 50 40 256 182, clip, angle=0, height=0.155\textheight] {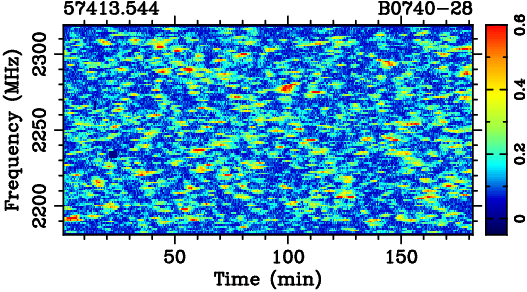} 
    \includegraphics[bb = 50 40 260 182, clip, angle=0, height=0.155\textheight] {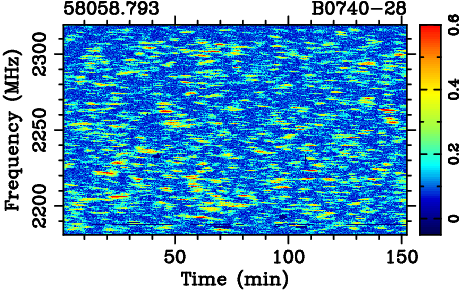} \\
    \includegraphics[bb = 32 32 279 272, clip, angle=0, height=0.154\textheight] {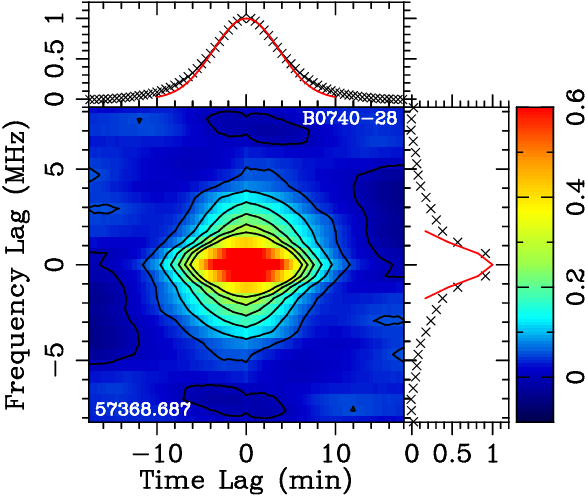}
    \includegraphics[bb = 65 32 279 272, clip, angle=0, height=0.154\textheight] {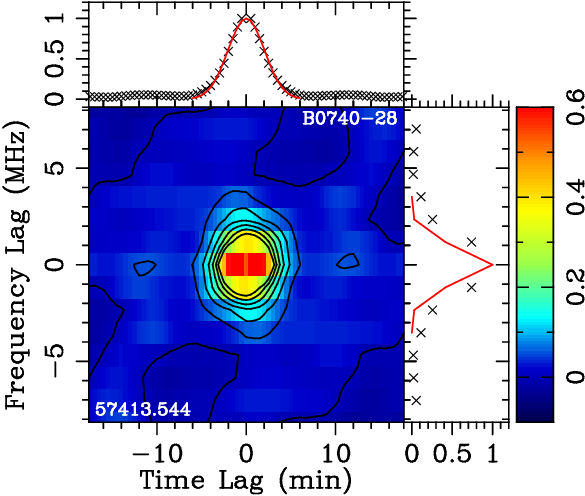}
    \includegraphics[bb = 65 32 320 272, clip, angle=0, height=0.154\textheight] {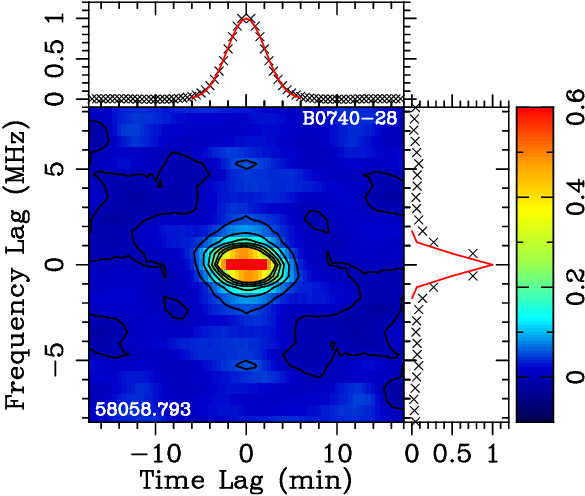}
  \caption{Dynamic spectra from three observations of PSR B0740-28
    ({\it upper panels}) and their autocorrelation functions ({\it lower
      panels}). See keys in Fig.~\ref{fig0329}. } 
  \label{fig0740} 
\end{figure*}


%
\begin{figure}
  \includegraphics[bb = 25 40 205 182, clip,angle=0, height=0.15\textheight] {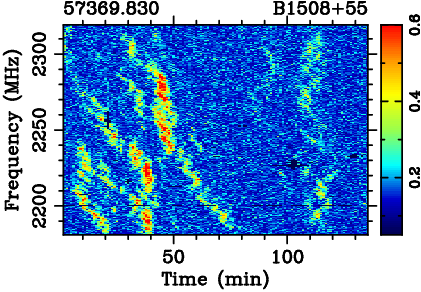}
  \includegraphics[bb = 50 40 152 182, clip,angle=0, height=0.15\textheight] {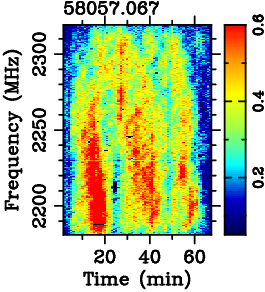} \\
  \includegraphics[bb = 32 32 279 272, clip,angle=0, height=0.15\textheight] {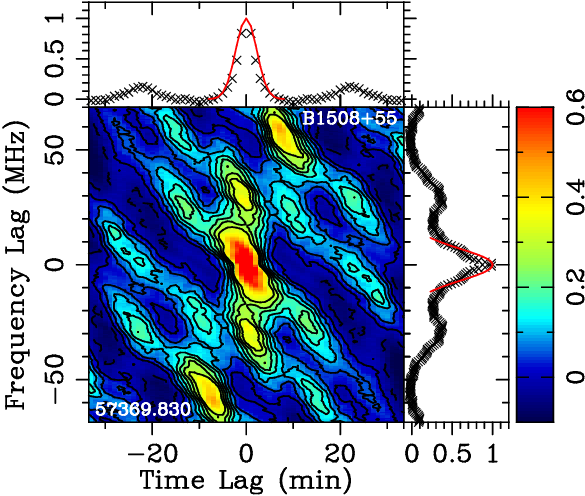}
  \includegraphics[bb = 65 32 320 272, clip,angle=0, height=0.15\textheight] {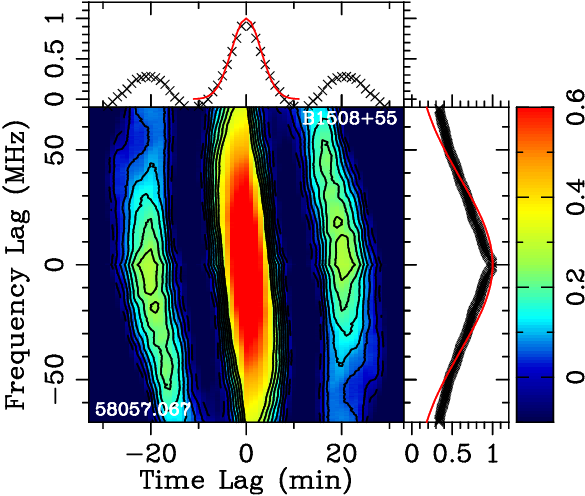} 
  \caption{Dynamic spectra from two observations of PSR B1508+55 ({\it
      upper two panels}) and their autocorrelation functions ({\it
      lower two panels}). See keys in Fig.~\ref{fig0329}. }
  \label{fig1508}
\end{figure}


\begin{figure*}
  \begin{tabular}{l}
    \includegraphics[bb = 25 40 93  172, clip, angle=0, height=0.155\textheight] {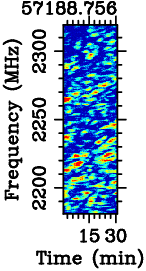}
    \includegraphics[bb = 50 40 332 172, clip, angle=0, height=0.155\textheight] {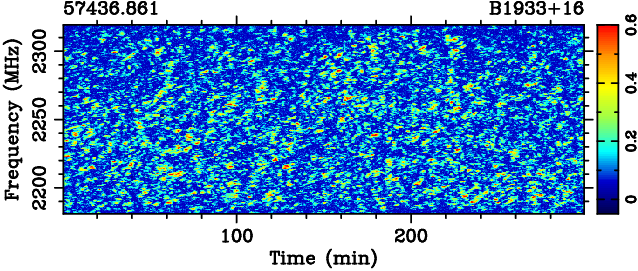} \\
    \includegraphics[bb = 25 40 527 172, clip, angle=0, height=0.155\textheight] {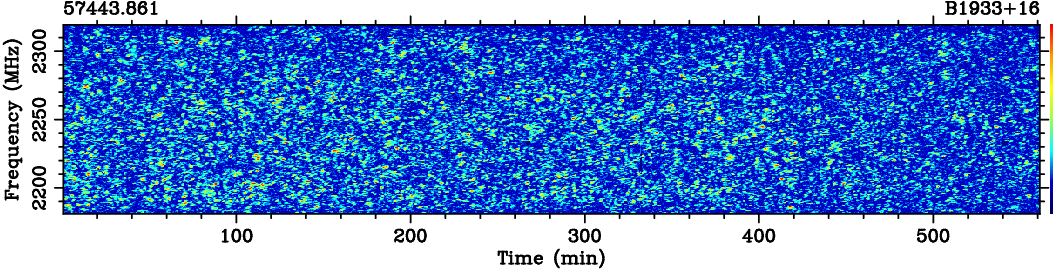} \\
    \includegraphics[bb = 25 40 370 172, clip, angle=0, height=0.155\textheight] {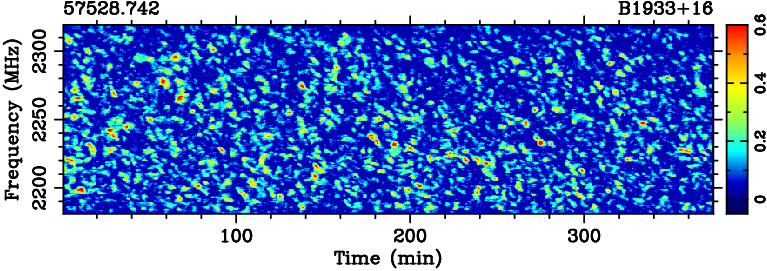} \\
    \includegraphics[bb = 25 40 505 172, clip, angle=0, height=0.155\textheight] {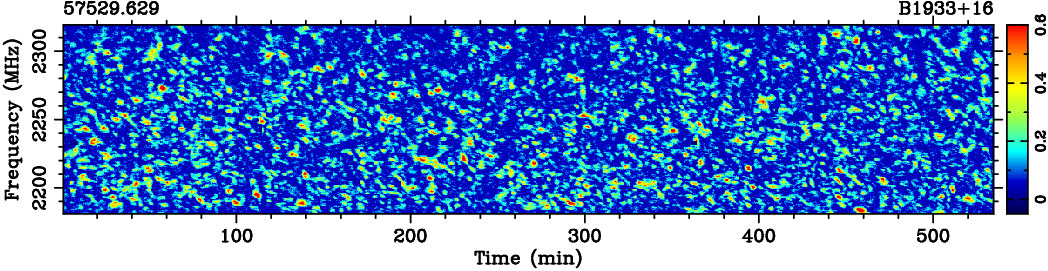} \\
    \includegraphics[bb = 25 40 505 172, clip, angle=0, height=0.155\textheight] {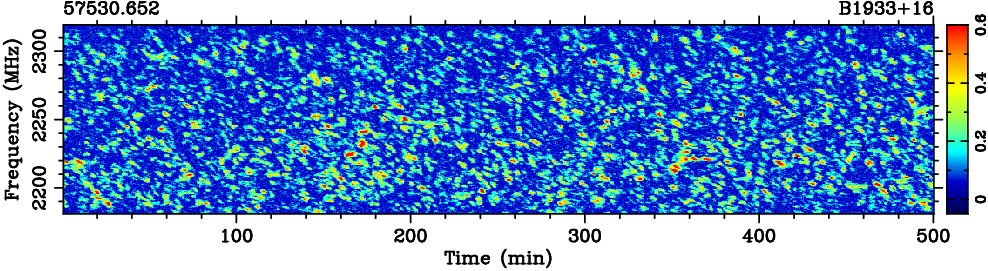} \\
    \includegraphics[bb = 25 40 385 172, clip, angle=0, height=0.155\textheight] {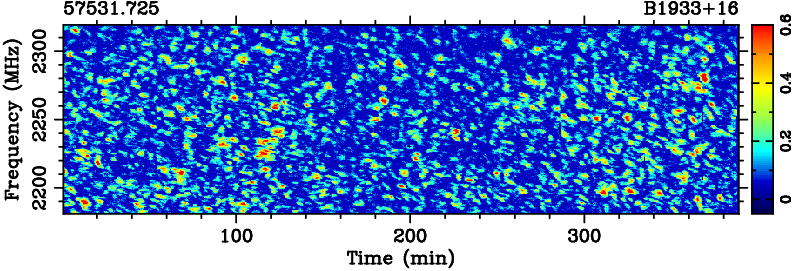}
    \includegraphics[bb = 50 40 225 172, clip, angle=0, height=0.155\textheight] {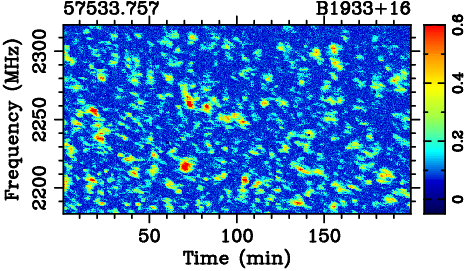}
  \end{tabular}
  \caption{Dynamic spectra from eight observations of PSR
    B1933+16. See keys in Fig.~\ref{fig0329}.}
  \label{fig1933DS}
\end{figure*}
\begin{figure*}
  \begin{tabular}{l}
    \includegraphics[bb = 32 62 279 272, clip, angle=0, height=0.131\textheight] {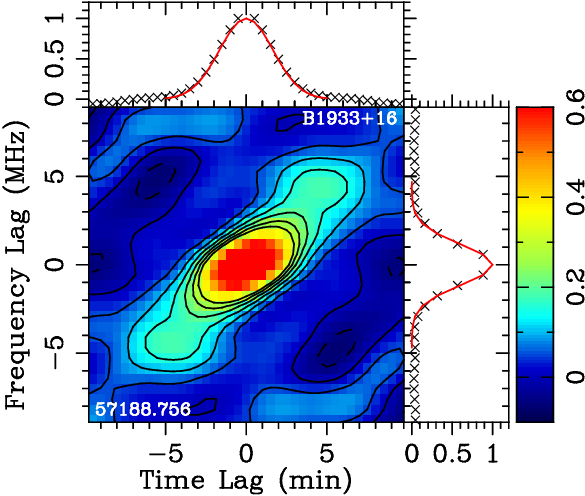}
    \includegraphics[bb = 67 62 279 272, clip, angle=0, height=0.131\textheight] {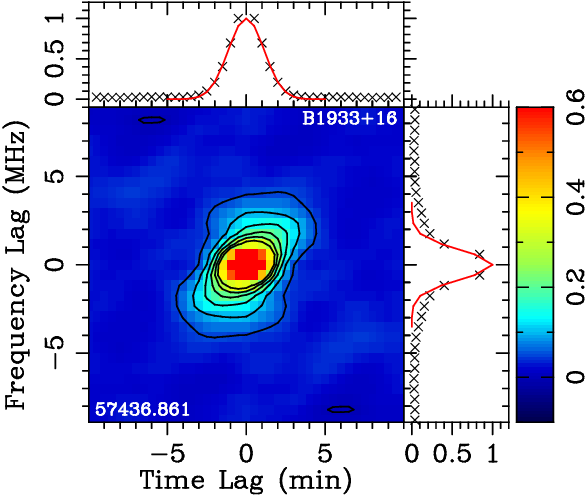}
    \includegraphics[bb = 67 62 279 272, clip, angle=0, height=0.131\textheight] {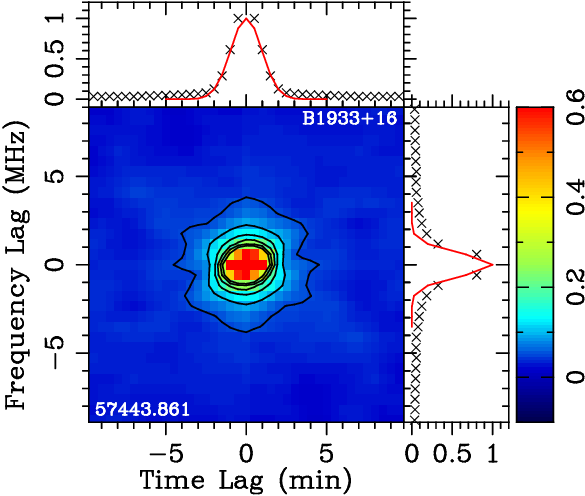}
    \includegraphics[bb = 67 62 320 272, clip, angle=0, height=0.131\textheight] {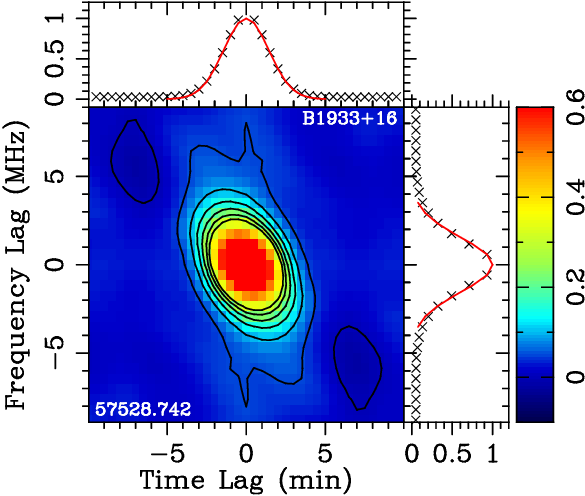} \\
    \includegraphics[bb = 32 32 279 272, clip, angle=0, height=0.15\textheight] {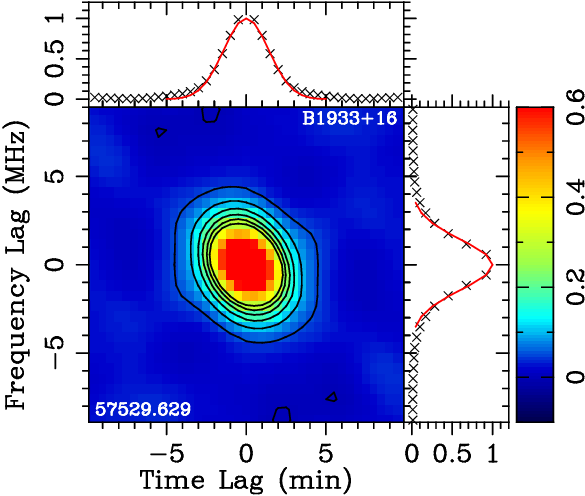}
    \includegraphics[bb = 67 32 279 272, clip, angle=0, height=0.15\textheight] {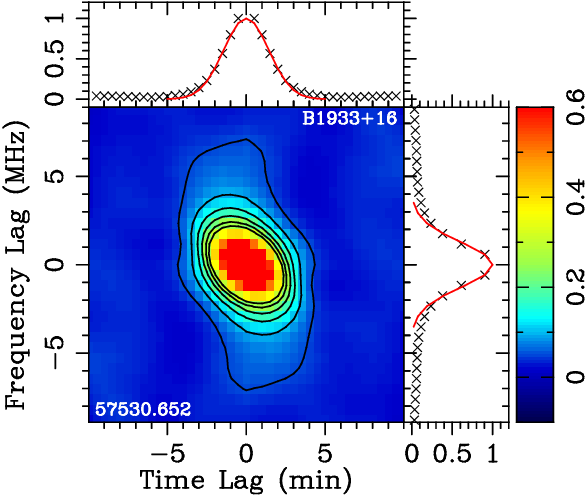}
    \includegraphics[bb = 67 32 279 272, clip, angle=0, height=0.15\textheight] {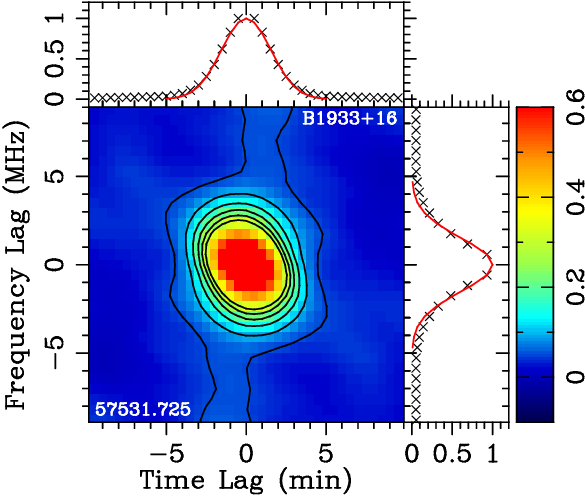}
    \includegraphics[bb = 67 32 279 272, clip, angle=0, height=0.15\textheight] {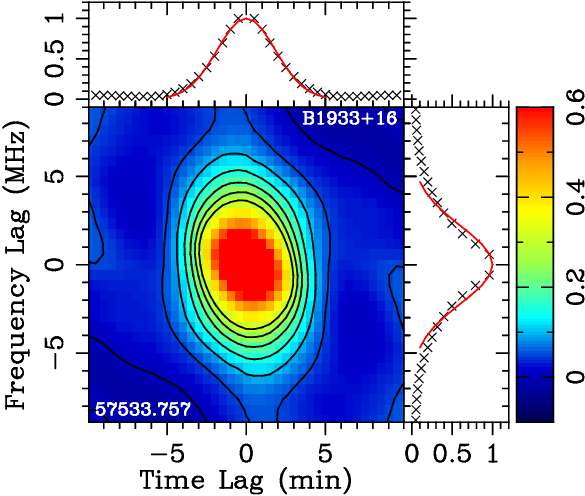}
  \end{tabular}
  \caption{The autocorrelation functions of dynamic spectra for eight
    observations of PSR B1933+16. See keys in Fig.~\ref{fig0329}.}
  \label{fig1933AC}
\end{figure*}


\begin{figure*}
  \begin{tabular}{l}
    \includegraphics[bb = 25 40 220 182, clip, angle=0, height=0.155\textheight] {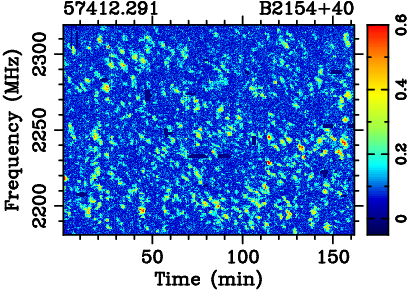} 
    \includegraphics[bb = 50 40 302 182, clip, angle=0, height=0.155\textheight] {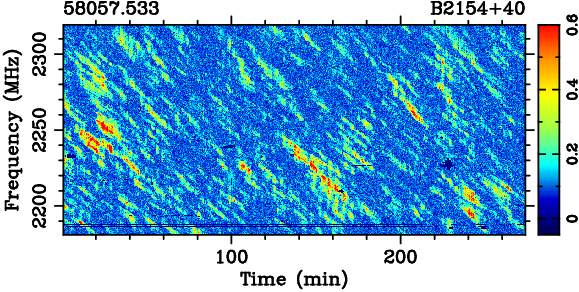}
    \includegraphics[bb = 50 40 290 182, clip, angle=0, height=0.155\textheight] {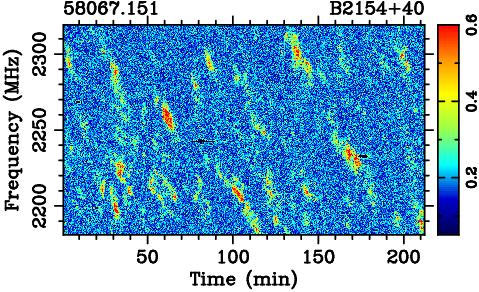} 
  \end{tabular}
  \begin{tabular}{c}
    \includegraphics[bb = 32 32 279 272, clip, angle=0, height=0.154\textheight] {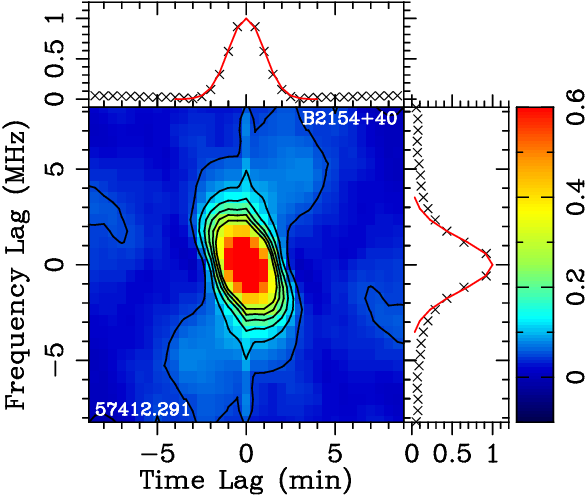}
    \includegraphics[bb = 64 32 279 272, clip, angle=0, height=0.154\textheight] {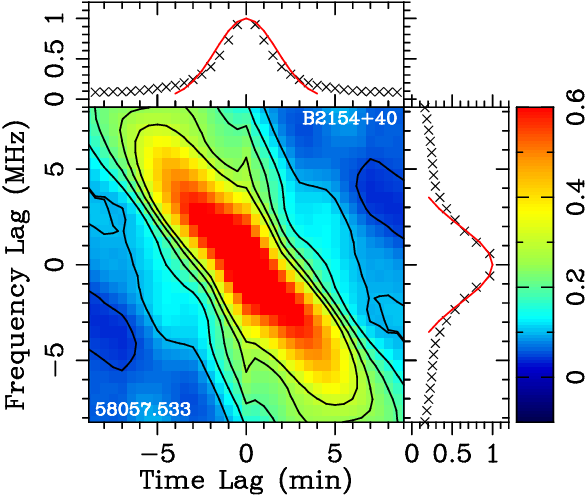}
    \includegraphics[bb = 64 32 320 272, clip, angle=0, height=0.154\textheight] {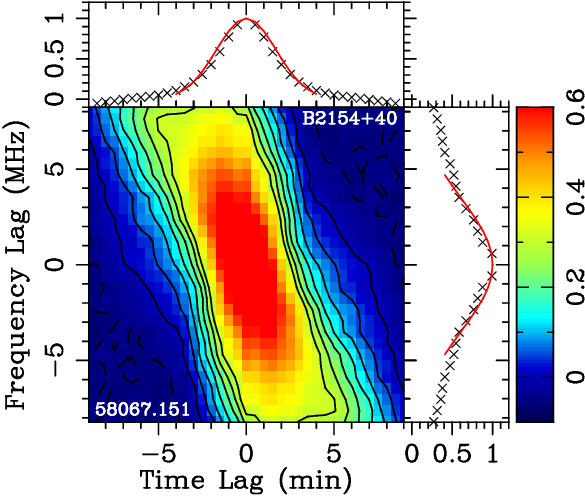}
  \end{tabular}
  \caption{Dynamic spectra ({\it upper three panels}) and
    autocorrelation functions ({\it bottom three panels}) for three
    observations of PSR B2154+40. See keys in Fig.~\ref{fig0329}.  }
  \label{fig2154}
\end{figure*}


\begin{figure}
  \begin{tabular}{l}
    \includegraphics[bb = 25 40 250 182, clip, angle=0, height=0.15\textheight] {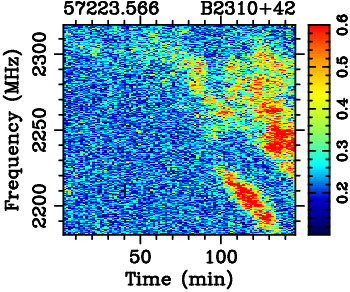} \\
    \includegraphics[bb = 25 40 400 182, clip, angle=0, height=0.15\textheight] {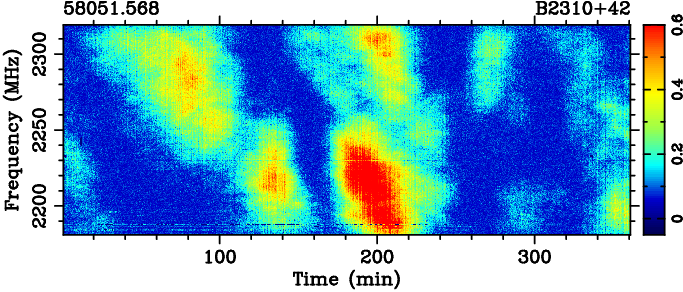} \\
    \includegraphics[bb = 32 32 279 272, clip, angle=0, height=0.14\textheight] {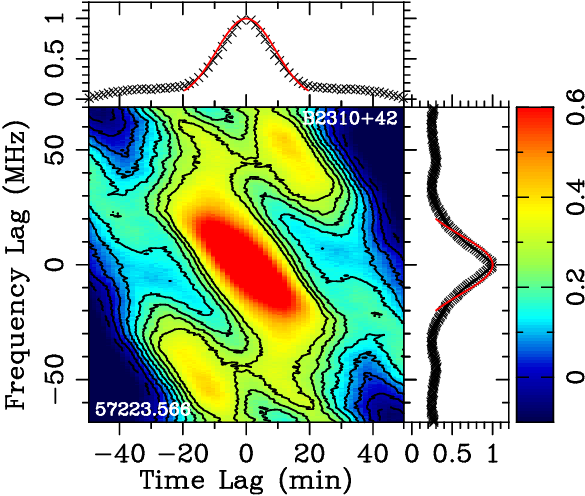}
    \includegraphics[bb = 65 32 320 272, clip, angle=0, height=0.14\textheight] {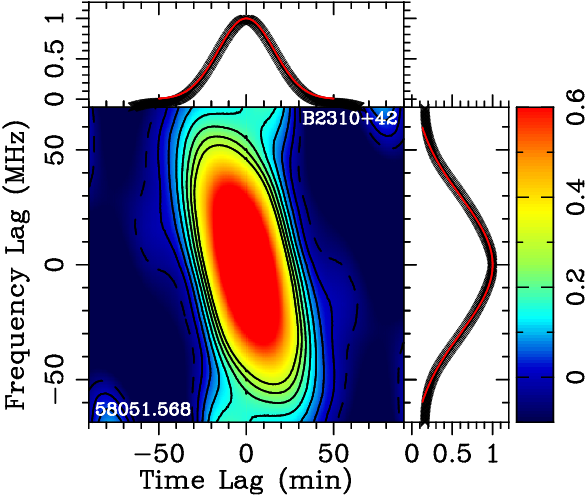}
  \end{tabular}
  \caption{Dynamic spectra from two observations of PSR B2310+42 ({\it
      upper two panels}) and the autocorrelation functions ({\it lower
      two panels}). See keys in Fig.~\ref{fig0329}. }
  \label{fig2310}
\end{figure}
\begin{figure}
  \begin{tabular}{l}
    \includegraphics[bb = 25 40 250 182, clip, angle=0, height=0.15\textheight] {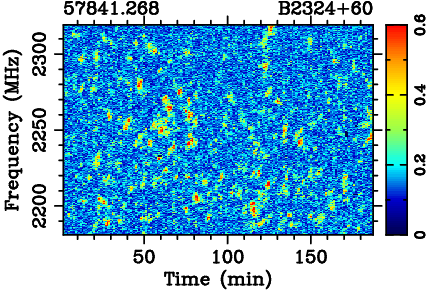} \\
    \includegraphics[bb = 25 40 400 182, clip, angle=0, height=0.15\textheight] {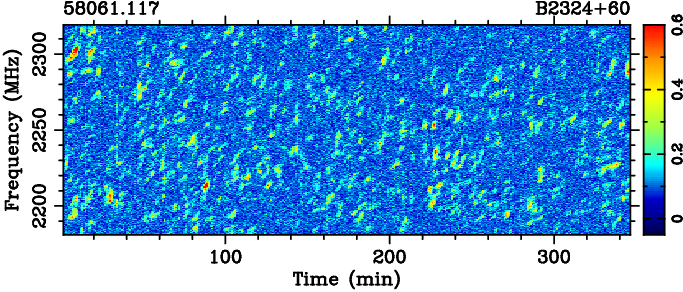} \\
    \includegraphics[bb = 32 32 279 272, clip, angle=0, height=0.14\textheight] {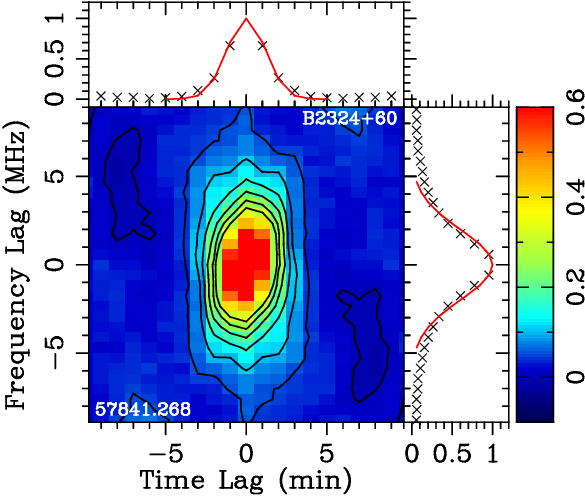}
    \includegraphics[bb = 64 32 320 272, clip, angle=0, height=0.14\textheight] {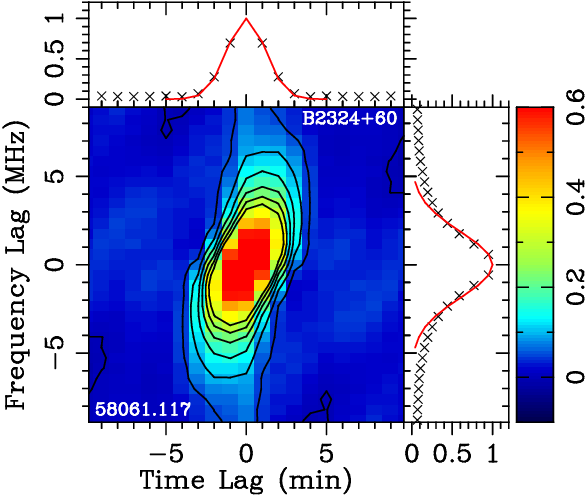}
  \end{tabular}
  \caption{Dynamic spectra from two observations of PSR B2324+60 ({\it
      upper two panels}) and the autocorrelation functions ({\it lower
      two panels}). See keys in Fig.~\ref{fig0329}. }
  \label{fig2324}
\end{figure}
\begin{figure}
  \begin{tabular}{l}
    \includegraphics[bb = 25 40 128 182, clip, angle=0, height=0.15\textheight] {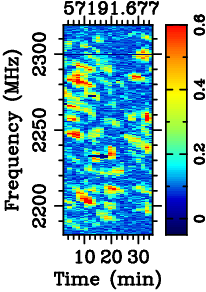}
    \includegraphics[bb = 51 40 200 182, clip, angle=0, height=0.15\textheight] {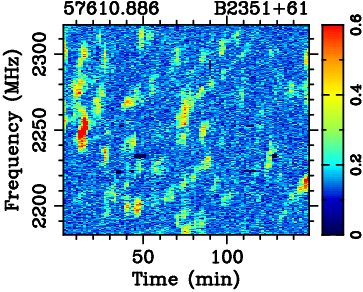} \\
    \includegraphics[bb = 32 32 279 272, clip, angle=0, height=0.14\textheight] {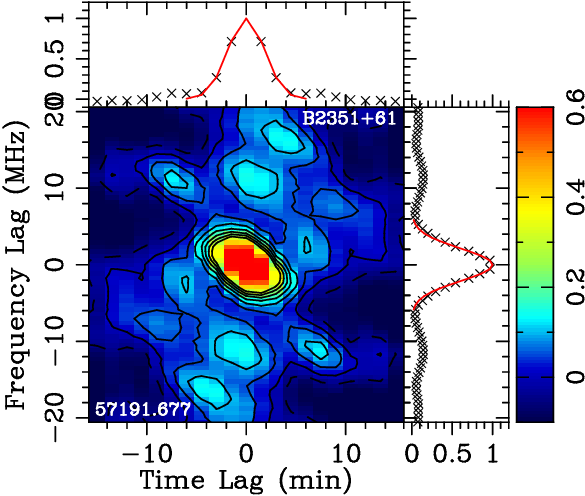}
    \includegraphics[bb = 64 32 320 272, clip, angle=0, height=0.14\textheight] {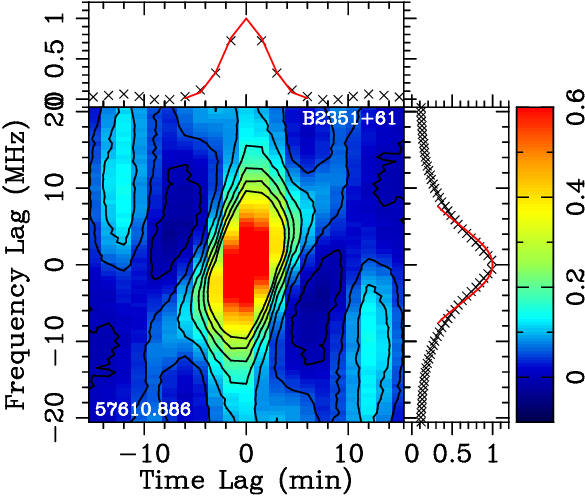}
  \end{tabular}
  \caption{Dynamic spectra from two observations of PSR B2351+61 ({\it
      upper panels}) and the autocorrelation functions ({\it lower
      panels}). See keys in Fig.~\ref{fig0329}.}
  \label{fig2351}
\end{figure}

\section{Dynamic spectra and derived scintillation parameters}

In the following, we discuss scintillation observations for each
pulsar.


\subsection{PSR B0329+54}

PSR B0329+54 is a bright pulsar in the northern sky. Its scintillation
has been observed by many authors (see Table~\ref{tab2}) in a range of
frequencies from 327~MHz to more than 10~GHz. Though the decorrelation
bandwidths and scintillation time scales have somehow different values
from many observations even at one given band \citep[e.g.][]{sfm96,
  wymw08}, the scintillation parameters clearly follow a power-law
with observational frequencies \citep{cwb85,lkgk11}, which will be
discussed in Sect.~\ref{sfd}.

We made three long observations at the S-band for 306, 426 and 655
minutes, respectively, and obtained their dynamic spectra (see
Fig.~\ref{fig0329}) which were at the highest frequency currently
available. Obviously, the scintles are of various scales in different
sessions. We carefully inspected the intensity variations of
individual pulses, and realized that the pulses were modulated and not
so stable within a subintegration time of less than 30~s. The
normalization for the subintegration by its own total power could
reduce the modulation, but might distort the dynamic spectra. Hence,
we got the mean pulse intensity for each channel with a subintegration
time of $\rm \delta t= 30$~s (or 60~s or 90~s for other pulsars, see
Table~\ref{tab3}) without normalization. The scintillation parameters
are then derived from the auto-correlation functions. The
scintillation time scale ranged from 19 to 30 minutes at the S-band
(see Table~\ref{tab3}).  Long observations are thus necessary to get
enough independent scintles in the dynamic spectra.


\subsection{PSR B0355+54}

PSR B0355+54 was previously observed for interstellar scintillation in
the frequency range from 325~MHz to more than 10~GHz (see
Table~\ref{tab2}), and dynamic spectra were published by
\citet{whs74}, \citet{sti07} and recently by \citet{xlh+18}.

We made five observations for this pulsar, one of which lasted for as
long as 720 minutes. The new observations showed quite different
dynamic spectra (see Fig.~\ref{fig0355}). The first observation at
57254 exhibited a wide-scintillation band and long time scale (see
Table~\ref{tab3}), so did the last two observations at 58062 and
58066. However, the two observations in the middle session at 57416
gave not only a much narrower scintillation band and finer time scale,
but also fringes periodic in both frequency and time, and even with
crossed fringes with different drifting rates in the dynamic
spectra. These features were well-shown in the secondary spectra and
would be further discussed in Sect.\ref{s2s}.

The scintillation parameters (see Table~\ref{tab3}) were also
estimated from the newly observed dynamic spectra. The decorrelation
bandwidth was at the highest frequency (see Table~\ref{tab2}), which
could be used to check the power-law behavior of fluctuations of the
interstellar medium, see Sect.\ref{sfd}. The substantial decrease in
the decorrelation bandwidth at the epoch of 57416 indicated a great
increase in the scintillation strength $u$, similar as the extreme
scattering event reported in \citet{kcw+18}.


\subsection{PSR B0540+23}

PSR B0540+23 was previously observed for interstellar scintillation
from 430~MHz to more than 10~GHz (see Table~\ref{tab2}), and only the
decorrelation bandwidth or only scintillation time scale was obtained
at a given frequency \citep{cwb85,mss+96}. Our observations provided
the first dynamic spectra (see Fig.~\ref{fig0540}), which gave the
decorrelation bandwidth at the highest frequency and the scintillation
time-scale at the lowest frequency. These new measurements in the
three sessions also showed significant variations of the scintle
sizes. Well-organized fringes can be marginally recognized, which is
reflected in the secondary spectrum as discussed in Sect.\ref{s2s}.

\subsection{PSR B0740$-$28}

PSR B0740$-$28 was previously observed for interstellar scintillation
by \citet{mss+96,jnk98} at 660~MHz, 4.75, 4.8, 8.4 and 10.55~GHz, and
for the scattering at lower frequencies by \citet{sod80}. We made
three observations for this pulsar, one with a channel width of
1.17~MHz and other two with 0.58~MHz. The dynamic spectra (see
Fig.~\ref{fig0740}) showed that the decorrelation bandwidths were
small and comparable to the channel width for the last two
observations (see Table~\ref{tab3}). The decorrelation bandwidths
obtained by our S-band observations were at the lowest frequency ever
published, and the scintillation time-scale varied from 3.0 to 5.2
minutes.

\subsection{PSR B1508+55}

PSR B1508+55 was previously observed for scintillation at frequencies
only below 1.0~GHz. Our S-band observations were carried out at the
highest frequency up to now. Dynamic spectra for two observations were
shown in Fig.~\ref{fig1508}. The decorrelation bandwidths and
scintillation time scales varied a lot even during observations. The
derived parameters are given in Table~\ref{tab3}.

\subsection{PSR B1933+16}

PSR B1933+16 was previously observed for interstellar scintillations
at frequencies from 1410 to 1670~MHz (see Table~\ref{tab2}). The
decorrelation bandwidths were estimated at all the frequencies, but
only one scintillation time scale was available at 1670~MHz.

We made eight observations, the longest of which lasted for 564
minutes. These observations were at the highest frequency available
and exhibited clearly resolved dynamic spectra shown as numerous small
scintles (see Fig.~\ref{fig1933DS}). The scintles did not change much
in each session, but did vary among observations as demonstrated by
auto-correlation functions in Fig.~\ref{fig1933AC}. The observation at
the epoch of 57443 showed the smallest scintles with decorrelation
bandwidth and scintillation time scale of 0.77~MHz and 1.37~min. The
relatively small decorrelation bandwidth was a bit larger than the
channel width (see Table~\ref{tab3}). The observation at the epoch of
57533 showed the largest scintles with scintillation parameters of
2.58~MHz and 2.61~min. The thin screen model clearly predicted the
variations of the scintillation parameters within the schemes of
diffractive and refractive scintillations, which will be discussed in
detail in Sect.~\ref{s1933}.

\subsection{PSR B2154+40}

PSR B2154+40 was previously observed by \citet{cor86} for interstellar
scintillation, whose decorrelation bandwidths and scintillation
time-scales were scaled to 1~GHz for the investigation of velocities
for an ensemble of pulsars. The recent observations by \citet{spg+17}
gave not only the dynamic spectra at 1420~MHz but also the parabolic
arcs of the secondary spectrum. They found that the scattering screen
was close to the pulsar at that time, and about 2.67 kpc away from the
observer (cf. the distance to the pulsar was taken to be 2.90~kpc).

Our three observations gave the dynamic spectra (see
Fig.~\ref{fig2154}). The scintillation parameters were derived (see
Table~\ref{tab3}). Apparently scintillation features varied a lot
among the observations. The three observations exhibited small
scintles, periodic fringes and sparsely separated scintles
respectively. The secondary spectrum of observation at the epoch of
58057 is shown in Sect.\ref{s2s}.

\subsection{PSR B2310+42}

PSR B2310+42 was previously observed for interstellar scintillation by
\citet{brg99a} at 327~MHz. Our S-band observations only detected
several scintles in the dynamic spectra (see Fig.~\ref{fig2310}) and
gave very large decorrelation bandwidths and time scales (see
Table~\ref{tab3}).

\subsection{PSR B2324+60}

No previous observations have ever been made for scintillation of PSR
B2324+60, and ours are the first. The dynamic spectra are shown in
Fig.~\ref{fig2324}. Apparently the scintles are very small, similar to
those of PSR B1933+16. Nevertheless the decorrelation bandwidths and
scintillation time-scales (see Table~\ref{tab3}) derived from the
dynamic spectra are just a few times of the resolutions of the
frequency channel and subintegration time.

\subsection{PSR B2351+61}

PSR B2351+61 was previously observed for interstellar scintillation by
\citet{mss+96} at 10.55~GHz, and they got the scintillation time
scale. No dynamic spectra have ever been obtained for this pulsar. Our
two observations provided the dynamic spectra (see Fig.~\ref{fig2351})
for the first time. Scintillation patterns vary a lot among the
observations with the later one exhibiting sparsely distributed
scintles, so that the derived decorrelation bandwidths and
scintillation time scales (see Table~\ref{tab3}) were quite different
in two sessions.

\section{Discussion}

The interstellar scintillation that we observed at S-band belongs to
the category of strong scattering, which is supported by the large
values of the derived scintillation strength $u$, much larger than the
critical value of 1.0 as listed in Table~\ref{tab3}. For $u>1$, the
scintillation breaks into two branches, refractive and diffractive
schemes. The average level of turbulence $C_n^2$ are estimated by
using Equation~\ref{eq:cn2}, which are generally consistent with
previous measurements \citep[e.g.][]{cwb85,wmj+05} though they vary
among different sessions.

In this section, the Kolmogorov feature of the interstellar turbulence
will be investigated by analyzing the frequency dependencies of the
scintillation parameters by combing our new observations with previous
observations of the ten pulsars. Moreover, pulsar velocities will also
be estimated from scintillation patterns and compared with those from
VLBI measurements. By using several long observations, we discuss the
diffractive and refractive scintillation of PSR B1933+16. The clear
fringes in the dynamic spectra will be investigated for three pulsars,
PSRs B0355+54, B0540+23 and B2154+40 through the secondary spectra.

\begin{figure}
  \centering
  \includegraphics[bb = 30 52 279 272, clip, angle=0, height = 0.16\textheight] {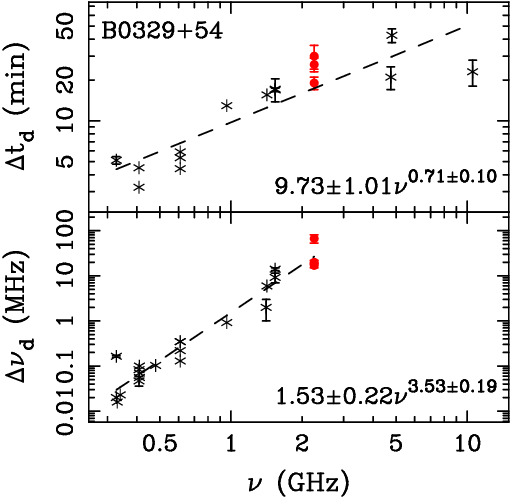}
  \includegraphics[bb = 56 52 279 272, clip, angle=0, height = 0.16\textheight] {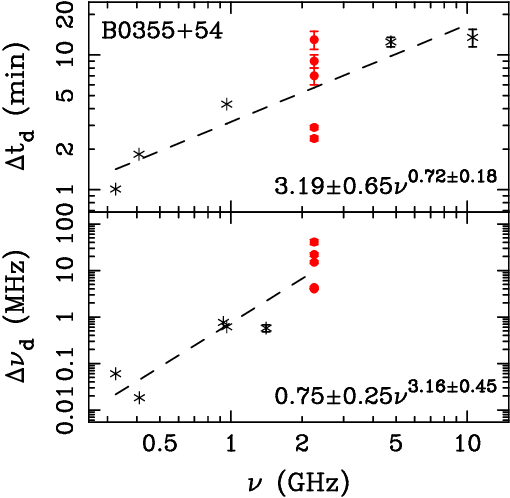}\\
  \includegraphics[bb = 30 52 279 272, clip, angle=0, height = 0.16\textheight] {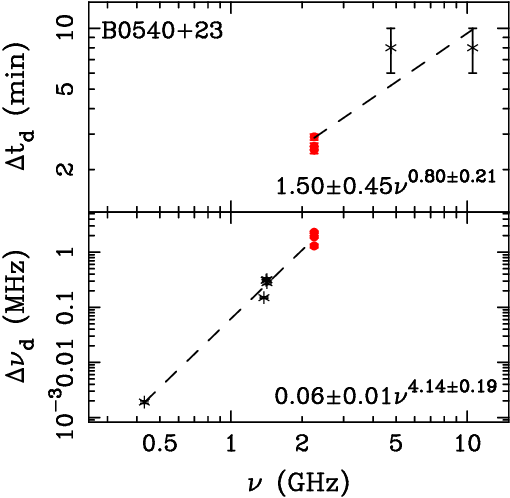}
  \includegraphics[bb = 56 52 279 272, clip, angle=0, height = 0.16\textheight] {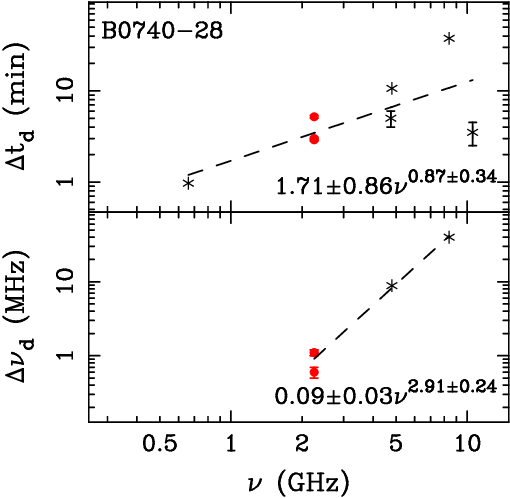}\\
  \includegraphics[bb = 30 52 279 272, clip, angle=0, height = 0.16\textheight] {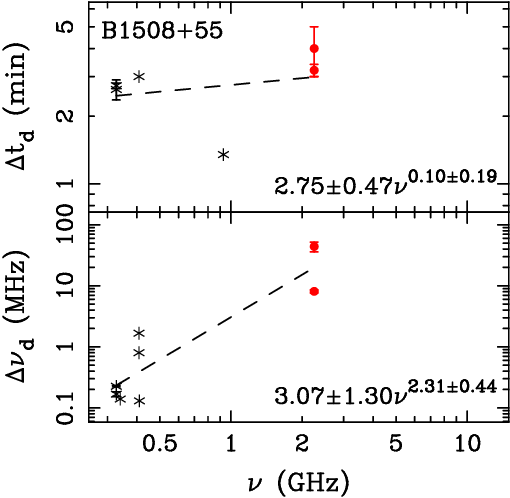}
  \includegraphics[bb = 56 52 279 272, clip, angle=0, height = 0.16\textheight] {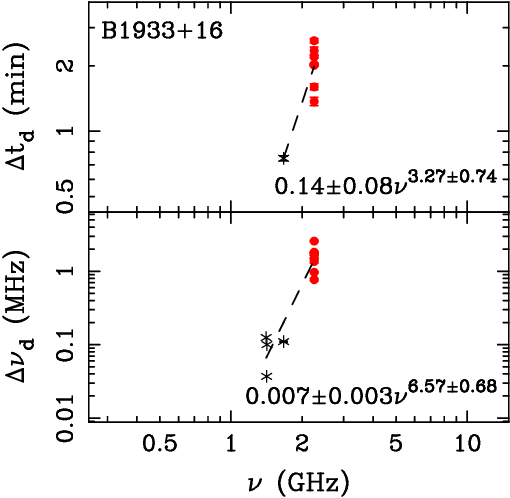}\\
  \includegraphics[bb = 30 33 279 272, clip, angle=0, height = 0.173\textheight] {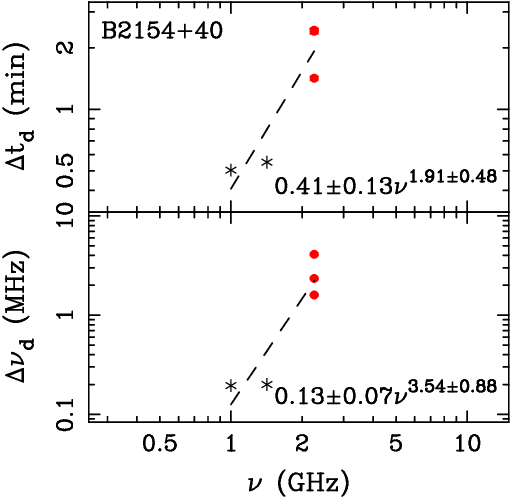}
  \includegraphics[bb = 56 33 279 272, clip, angle=0, height = 0.173\textheight] {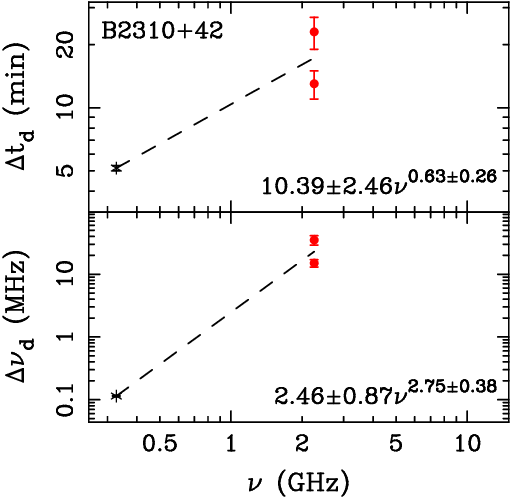}\\
  \caption{Frequency dependencies of decorrelation bandwidths and
    scintillation time-scales for eight pulsars. Asterisks represent
    previous measurements listed in Table~\ref{tab2}, and dots are our
    new measurements as listed in Table~\ref{tab3}. The dashed lines
    represent the best power-law fitting as given in the lower-right
    corner of each panel.}
  \label{specAll}
\end{figure}

\subsection{Frequency dependence of scintillation parameters}
 \label{sfd}

Observations with a wide frequency range are important to investigate
the frequency dependencies of scintillation parameters and then the
turbulent feature of the interstellar medium.  For the widely accepted
Kolmogorov turbulence, it was predicted that $\Delta t_d \propto
\nu^{\alpha_t=6/5}$, $\Delta \nu_d \propto \nu^{\alpha_{\nu}=22/5}$,
as shown by Equation~\ref{eq:parafreq}. Combining previous
measurements at various frequencies in Table~\ref{tab2} with our newly
determined decorrelation bandwidths and scintillation time scales at
2250~MHz in Table~\ref{tab3}, we can investigate the frequency
dependence of scintillation parameters. Data are shown in
Fig.~\ref{specAll}, except for PSRs B2324+60 and B2351+61 because
merely our measurements at S-band are available and there are no data
at other frequencies. For seven pulsars our new data of decorrelation
bandwidths are at the highest frequencies, and for PSR B0740-28 ours
are at the lowest one. For four pulsars, PSRs B1508+55, B1933+16,
B2154+40 and B2310+42, only few measurements are available and our
measurements are crucial in determining the frequency dependencies of
their scintillation parameters. For PSRs B0329+54, B0355+54, B0540+23
and B0740$-$28, the scintillation parameters roughly agree with the
previous measurements by following a power-law. In general for these 8
pulsars, the power-indices for the decorrelation bandwidth vary
greatly from 2.91 to 4.23, and only that of PSR B0540+23 is close to
4.4 for the Kolmogorov turbulence. Therefore, power-indices for the
decorrelation bandwidth are generally smaller than the predictions by
the Kolmogorov turbulence. The power-indices for the scintillation
time scale vary greatly in the range from 0.71 to 0.87, also smaller
than the predictions by the Kolmogorov turbulence, which has
previously been noticed by
\citet{jnk98,bcc+04,wmj+05,lkgk11}. Moreover, by investigating the
characteristic time of scattering, \citet{gkk+17} also noticed that
the frequency dependence is more complex than simple power
law. Therefore the intervening medium between a pulsar and us can
rarely be ideally turbulent with a Kolmogorov spectrum.

It should be noted, however, that the interstellar turbulence should
be examined by using the scintillation parameters at different
frequencies observed at the same epoch for the same medium, because
pulsars move fast and the interstellar medium that they shine through
is different at different epochs. Ideal observations should be done
with a wide-band receiver to determine the instantaneous turbulence
feature of the interstellar medium.

\subsection{Scintillation velocity}

The scintillation velocity $V_{\rm s}$ is effectively the combination
of the transverse velocities of the pulsar, the intervening medium and
the Earth, which can be derived from scintillation observations
according to Equation~\ref{eq:viss}. The dominant contribution in
scintillation velocity is the transverse velocity of a pulsar, so that
the pulsar velocities measured by other means are well-correlated with
scintillation velocity \citep{ls82, cor86, gup95, brg99a}.

\begin{figure}
  \centering
   \includegraphics[width = 0.35\textwidth] {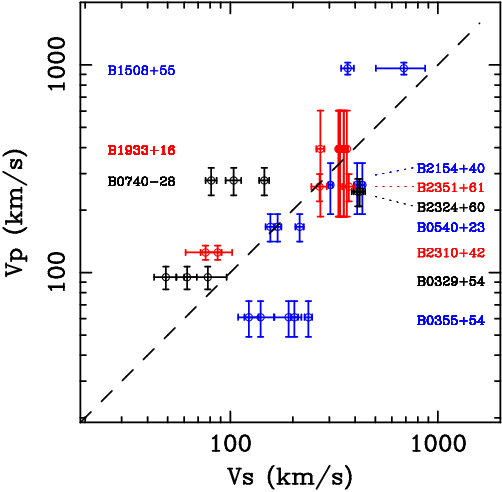}
  \caption{Comparison of proper motion velocities $V_{\rm p}$ (see
    Table~\ref{tab1}) of pulsars with scintillation velocities $V_{\rm
      s}$ (see Table~\ref{tab3}). The dashed line indicates the
    equality.}
  \label{vvpm}
\end{figure}

For ten pulsars that we observed (see Fig.~\ref{vvpm}), we estimated
$V_{\rm s}$ from our observations by using Equation~\ref{eq:viss}, as
listed in Table~\ref{tab3}, slightly different for different sessions
for a given pulsar. We find that scintillation velocities are roughly
consistent with proper motion velocities within two times the
uncertainties except for PSRs B0355+54, B0740-28 and B2324+60.

It is understandable for the velocity estimated by 
Equation~\ref{eq:viss} for a thin screen model to slightly deviate
from real velocities. \citet{gup95} attributed the differences for PSR
B0355+54 to the enhanced scattering at low heights above the Galactic
plane. As derived from its secondary spectrum analysis (see below and
Table~\ref{tab:sec}), the scattering screen of PSR B0355+54 locates
closely to the pulsar, while the velocity estimation based on
Equation~\ref{eq:viss} assumes that the screen lies at the midway
between the pulsar and us, which therefore is a significantly
over-estimate.  So is that of PSR B2154+40. Moreover, the intervening
medium rarely follows the Kolmogorov turbulence, as discussed above.

\begin{figure}
  \centering
  \includegraphics[width = 0.48\textwidth] {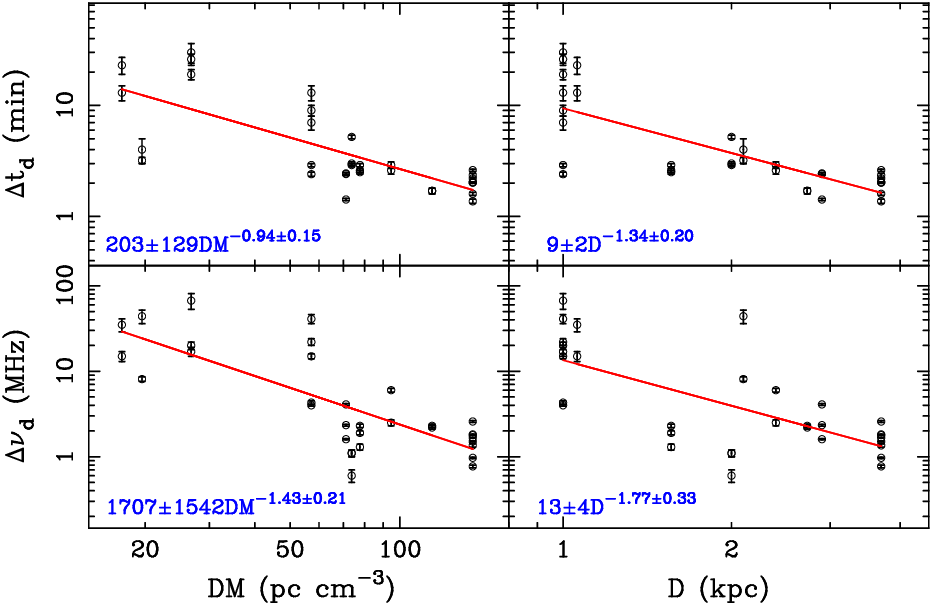}
  \caption{Scintillation parameters at 2250~MHz versus distance and
    dispersion measures for ten pulsars. The lines represent the best
    power-law fitting with the results given in the lower-left corner
    of each panel.}
  \label{DMAll}
\end{figure}

\subsection{Scintillation parameters versus distance}

For the Kolmogorov turbulence, scintillation parameters are related to
pulsar distance theoretically by $\Delta t_d \propto D^{-3/5}$,
$\Delta \nu_d \propto D^{-11/5}$, as shown by
Equation~\ref{eq:parafreq}. The distances are closely related to
dispersion measures of these pulsars. By analyzing a series of
measurements at 430~MHz, \citet{cwb85} found that $\Delta \nu_d
\propto \rm DM^{-\gamma}$ with power index $\gamma=3-4$, much steeper
than the theoretical prediction of $11/5$. This confirmed the previous
analysis by multi-frequency observations scaled to 408~MHz
\citep{ric70,ric77}, and it was attributed to variations of the level
of turbulence $C_n^2$ over a wide range of length scales.

The pulsars we observed cover a wider range of DMs (see
Table~\ref{tab1}) than those in literature. As clearly seen from
Figs.~\ref{fig0329} to \ref{fig2351}, pulsars with larger DMs tend to
have smaller scintles in the dynamic spectra. Fig.~\ref{DMAll} shows
the change of scintillation parameters with pulsar DM and distance.
Both $\Delta \nu_d$ and $\Delta t_d$ measured at S-band decrease with
DM and distance by following a power-law, with indices of
$-0.94\pm0.15$ and $-1.34\pm0.20$ for $\Delta t_d$ and $-1.43\pm0.21$
and $-1.77\pm0.33$ for $\Delta \nu_d$. The distance dependence for
$\Delta \nu_d$ is roughly consistent with the prediction by the
Kolmogorov turbulence with a large scatter of data, but the index for
$\Delta t_d$ is steeper than that of the Kolmogorov turbulence.  This
contrasts to results at low frequencies
\citep[e.g.][]{ric70,ric77,cwb85}.

\begin{figure}
  \centering
   \includegraphics[width = 0.35\textwidth] {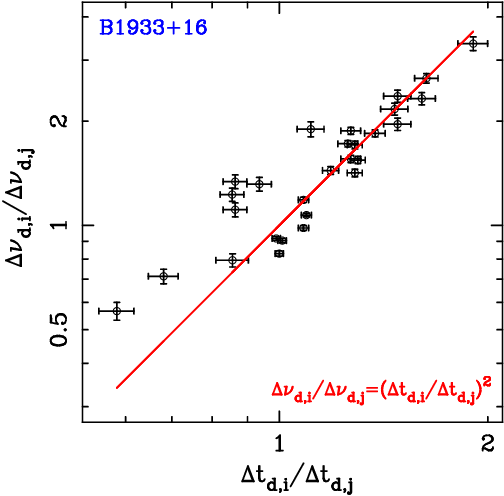}
   \caption{The ratio data of PSR B1933+16 for 28 paired epochs are
     consistent with the thin screen model which suggests that the
     ratio of decorrelation bandwidths of a paired epochs should have
     a quadratic relation with the ratio of scintillation time scales.
     A log-log plot is shown here with a linear fitting indicated by
     the solid line.}
  \label{dnuvsdt}
\end{figure}

\begin{figure}
  \centering
   \includegraphics[width = 0.35\textwidth] {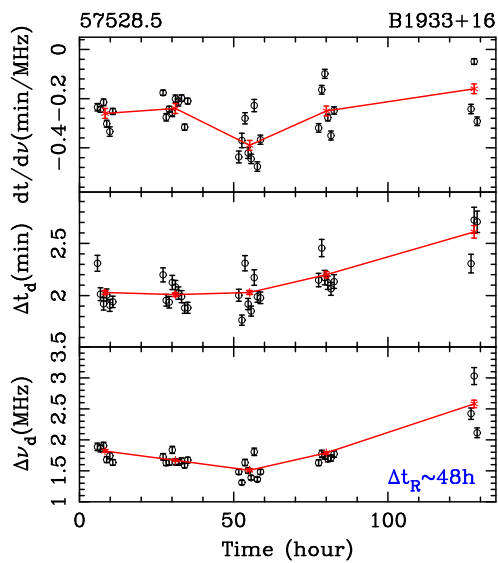}
   \caption{The scintillation parameters (open circles) calculated
     from every 1-hour block in 5 observational sessions of PSR
     B1933+16 since the epoch at 57528.5. They vary from those (dots)
     derived from data of an entire session listed in
     Table~\ref{tab2}. The refractive scintillation is indicated by
     the long term variations of these parameters at different
     sessions.}
  \label{scipara-evol}
\end{figure}

\subsection{Diffractive and refractive scintillation of PSR B1933+16}
\label{s1933}


In the diffractive scintillation scheme, the scatter broadened image
of a pulsar extends to an angular radius of $\theta_d$. In the thin
screen model \citep[e.g.][]{cor86, ric90}, it was predicted that
$\Delta \nu_d = C_1 c/(2 \pi D \theta_d^2)$ and $\Delta t_d=c/(2 \pi
\nu V_{\rm s} \theta_d)$. For a given pulsar, although scintillation
parameters vary among different observation sessions, both the ratio
of $\Delta \nu_d$ and the radio of $\Delta t_d$ from a paired sessions
should be related to $\theta_d^2$ or $\theta_d$, and therefore we get
\begin{equation}
  \Delta \nu_{d,i}/\Delta \nu_{d,j}=(\Delta t_{d,i}/\Delta t_{d,j})^2.
\end{equation}

By using the scintillation parameters ($\Delta \nu_d$, $\Delta t_d$)
obtained for the 8 observational sessions of PSR B1933+16 as listed in
Table~\ref{tab3}, we verified the above scaling relations, as shown in
Fig.~\ref{dnuvsdt}. For any pair of observations, $i$ and $j$, we get
$\Delta \nu_{d,i}/\Delta \nu_{d,j}$ and $\Delta t_{d,i}/\Delta
t_{d,j}$, and they follow the relation very well. We therefore
conclude that the thin screen model works well at S-band for the
pulsar scintillation.

Refractive scintillation accounts for focusing of rays within the
scattering disk with a scale $D \theta_d$. The time scale of
refractive scintillation is related to diffractive scintillation
parameters by \citep{sc90}
\begin{equation}
  \Delta t_R\simeq (4/\pi)(\nu/\Delta \nu_d) \Delta t_d.
\end{equation}
The scintillation parameters of PSR B1933+16 in Table~\ref{tab3} give
$\Delta t_R \sim 48$ hours, which is shorter than but comparable with
our quasi continuous observations of 130 hours since the epoch of
57528.5. To investigate their variations in detail, we obtained the
decorrelation bandwidth, scintillation time scale and the drift rate
for every 1-hour block of data, as shown in
Fig.~\ref{scipara-evol}. The variations at shorter time scales of
hours might be caused by diffractive scintillation, and obvious
long-term variations should come from the refraction from
irregularities of interstellar medium. However, it is hard to get
$\Delta t_R$ directly from the pulsar intensity fluctuations of our
limited data.

\begin{figure}
  \includegraphics[bb = 32 62 330 270, clip, angle=0, height=0.13\textheight] {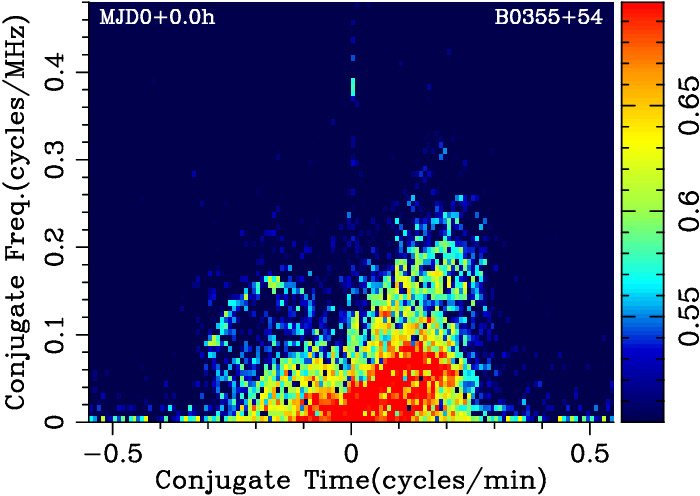}
  \includegraphics[bb = 72 62 330 270, clip, angle=0, height=0.13\textheight] {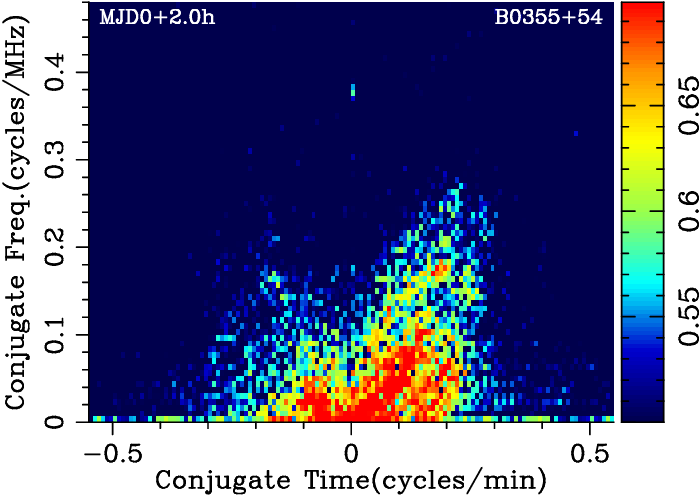}\\
  \includegraphics[bb = 32 62 330 270, clip, angle=0, height=0.13\textheight] {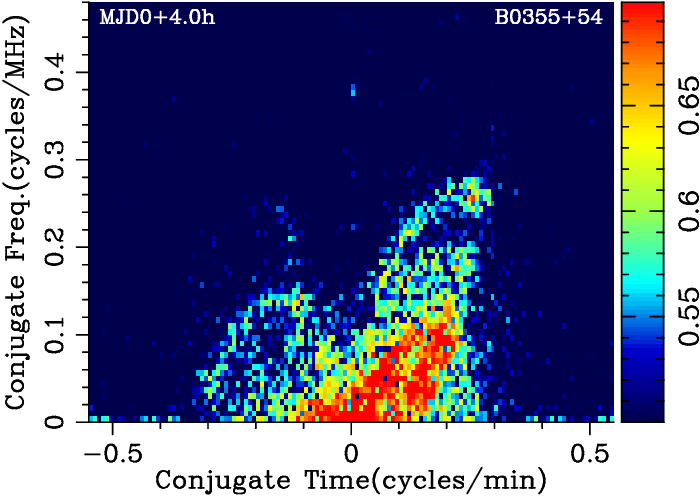}
  \includegraphics[bb = 72 62 330 270, clip, angle=0, height=0.13\textheight] {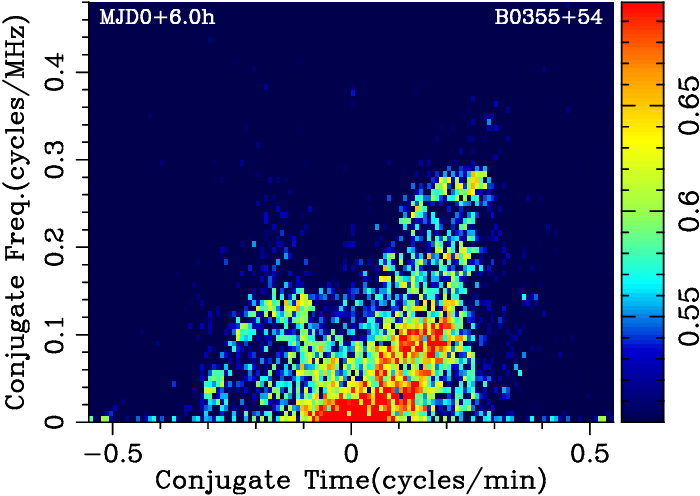}\\
  \includegraphics[bb = 32 32 370 270, clip, angle=0, height=0.148\textheight] {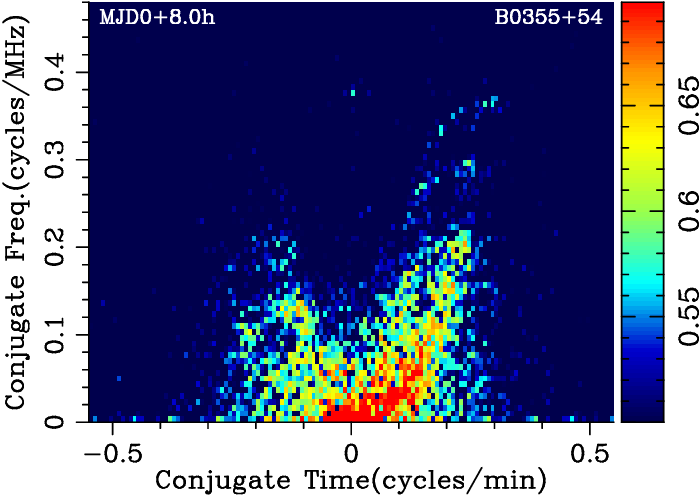}\\
  \includegraphics[width = 0.47\textwidth] {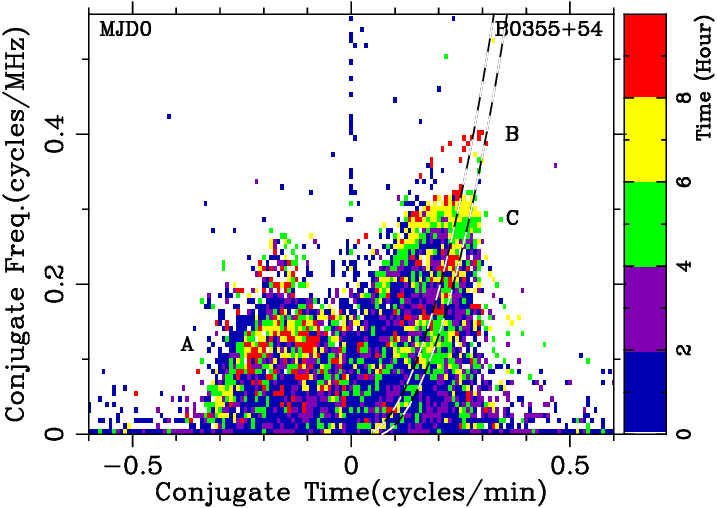}
  \caption{{\it Upper five panels:} Secondary spectra of PSR B0355+54
    for five 2-hour blocks of data started at MJD0 = 57416.036 + 105
    minutes, extracted from the 12 hour observation session of
    57416.036 (see Fig.~\ref{fig0355}). {\it The bottom panel}:
    Evolving secondary spectra plotted only for the maximum power
    among the five 2-hour blocks of data for each pixel. Different
    color scales are used to mark which block of data for the maximum
    power. Three special inverted arclets are labeled by A, B and C,
    and the main parabola is indicated by the dashed lines.}
  \label{ss0355}
\end{figure}

\begin{figure}
  \centering
  \includegraphics[bb = 32 32 330 270, clip, angle=0, height=0.143\textheight] {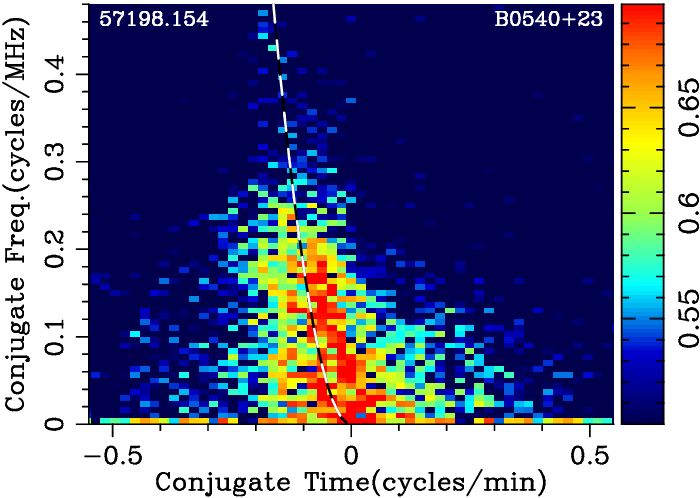}
  \includegraphics[bb = 72 32 370 270, clip, angle=0, height=0.143\textheight] {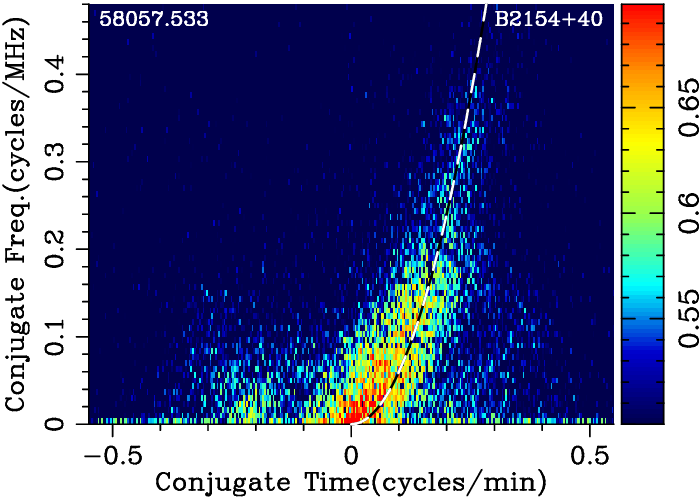}
  \caption{Secondary spectra of PSR B0540+23 ({\it left}) and
    PSR B2154+40 ({\it right}). }
  \label{ss-2}
\end{figure}


\begin{figure}
  \centering
  \includegraphics[width=0.45\textwidth] {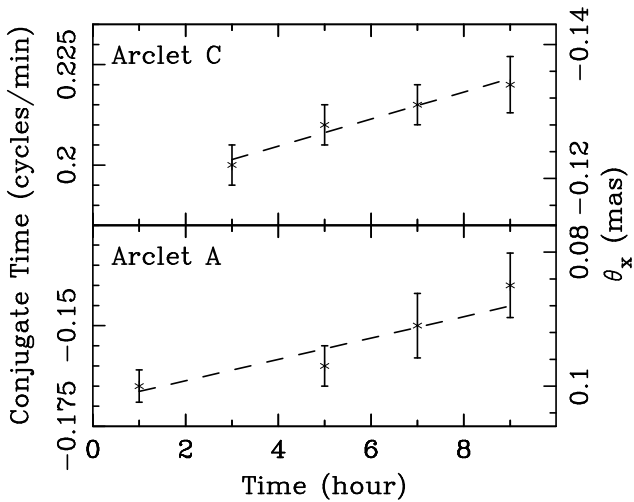}
  \caption{Variation of conjugate-time, $f_t$, of the vertexes of
    arclets A and C over a 12-hour session for PSR B0355+54. Their
    corresponding angular positions, $\theta_x$, can be found according
    to the scales in the right side of each panel.}
  \label{ss0355-evol}
\end{figure}

\subsection{Secondary Spectra of PSRs B0355+54, B0540+23 and B2154+40}
\label{s2s}

Well defined parabolic arcs have been detected in the secondary
spectra of dynamic spectra for some of our observational sessions of
PSRs B0355+54, B0540+23 and B2154+40, as shown in Fig.~\ref{ss0355}
and Fig.~\ref{ss-2}. They are very asymmetric about the conjugate
frequency axis. Previous detections of the arcs in the secondary
spectra were made by \citet{sti07} and \citet{xlh+18} for PSRs
B0355+54 and by \citet{spg+17} for PSR B2154+40. For PSR B0540+23,
this is the first time to report the detection of the arc, and its
left branch is much more significantly shown. For PSRs B0355+54 and
B2154+40, the right branches are much more significant. Refraction and
the corresponding relatively large phase gradients of the interstellar
medium are the main cause of the observed
asymmetry. An-extreme-scattering like event happened along the sight
line of PSR B0355+54 around MJD 57416, as indicated by the significant
decrease in the scintillation parameters and the sharply peaked arcs
of the secondary spectrum.

\begin{table}
  \centering
  \caption{The curvatures of parabolic arc, $\Omega$, of secondary
    spectra are used to estimate locations $d$ of scattering screens
    by assuming that pulsar velocity dominates and that any linear
    structure in the scattered image is aligned along the effective
    velocity vector. }
  \label{tab:sec}
  \tabcolsep 1mm
  \begin{tabular}{ccccccc}
    \hline
    \hline
PSRs & MJD & $\Omega$ &$ d $&$D_{\rm scr}$ &$r_v$&$d_{\rm iss}$ \\
         &   & ($s^3$)   &          & (kpc)      &       &     \\
    \hline
B0355+54 & 57416.036 & 0.0216 & 0.081 & 0.92 & 0.256 & 0.062 \\
B0540+23 & 57198.154 & 0.0648 & 0.556 & 0.69 & 0.982 & 0.491 \\
B2154+40 & 58057.533 & 0.0252 & 0.398 & 1.75 & 0.868 & 0.430 \\
   \hline
  \end{tabular}
\end{table}

The curvatures of the parabolic arcs, $\Omega$, can be estimated from
the fittings to secondary spectra, which are closely related to the
locations of the scattering screens $d$ \citep{hsb+03}, see the
derived parameters listed in Table~\ref{tab:sec} for three
pulsars. For PSR B0355+54, the curvature of the right part of the main
parabola in the secondary spectrum as indicated by the dashed lines in
Fig.~\ref{ss0355} has been used to estimate $d=0.081$. Based on the
known distances of pulsars in Table~\ref{tab1}, the real distances of
the screens $D_{\rm scr}$ are then determined. It should be noted that
Equation~\ref{eq:secspec} holds true only when the transverse
velocities for the observer ${\vec V}_{\rm obs}$ and the screen ${\vec
  V}_{\rm scr}$ are negligible in the effective velocity of ${\vec
  V}_{\rm eff}=(1-d) {\vec V}_{\rm p}+ d {\vec V}_{\rm obs}-{\vec
  V}_{\rm scr}$ and when the angle between the orientation of the
scattered image and ${\vec V}_{\rm eff}$ is small. In the thin screen
model, the location of the scattering screen can also be derived by
using $ r_v=V_{\rm p} / V_{\rm s} $ as introduced by \citet{gup95},
and the location of a screen is  \citep{smc+01}  $d_{\rm iss} =
r_v^2/(1+r_v^2)$. For PSR B0355+54, $d_{\rm iss}=0.062$ that is
consistent with $d$. We noticed that the screen we observed was
closer to the pulsar PSR B2154+40, very different from that observed
by \citet{spg+17}. The screen we observed for PSR B0540+23 was somehow
located halfway to the pulsar.

PSR B0355+54 also exhibits numerous isolated inverted parabolas in
Fig.~\ref{ss0355}, and their vertices vaguely presented near the
primary arc. The isolated arclets are probably caused by interference
of the fine substructures in the scattered images of the pulsar
\citep{crsc06}. These inverted arclets are also not distributed
symmetrically, and are located at different positions in the secondary
spectrum at different epochs. The arclet ``A'' in the left branch
gradually shifts downwards along the main parabola, and the arclet
``C'' in the right branch moves to right upwards, and the arclet ``B''
is detected only once at the latest epoch in the top right part of the
main parabola. This is only the second case to our knowledge that such
an evolution behavior of inverted arclets has been reported, and the
previous report was made for PSR B0834+06 by \citet{hsa+05} observed
over 26 days at 321 MHz.

The evolution of the arclets is more clearly shown in the bottom panel
of Fig.~\ref{ss0355} by plotting the maximum of each pixels of 5
blocks with different color scales. In order to quantitatively
describe their evolution, we determined the positions of the arclets
from each block of data by following \citet{hsa+05}. An inverted
arclet model was first constructed, which has similar curvature as
those in the secondary spectrum. The apex of the model was then fixed
to a given position in the secondary spectrum, and the power was
integrated along the arclet model. The two dimensional distribution of
arclet power across apex positions was calculated by sliding the model
across the secondary-spectrum plan. Through the plot, accurate
positions for arclets A and C were determined, which evolved with time
as shown in Fig.~\ref{ss0355-evol}.
Because the arclets are generally caused by separated deflecting
interstellar medium structures at various angles $\theta$, the
conjugate time $f_t$ (i.e. the abscissa of Fig.~\ref{ss0355}) can be
related to the angular position $\theta$ by $f_t=-\theta_x V_{\rm
  eff}/( \lambda d)$, here, $\lambda$ is the observing wavelength, and
$\theta_x$ is the angular position of the scattering structure along
the effective velocity direction. By measuring the moving rate of the
arclets across the main parabola, i.e., $\delta f_t/\delta t $, the
angular velocity for the detected scattering structures can be
estimated as
\begin{equation}
\delta \theta_x/\delta t= -(\lambda d/V_{\rm eff}) \delta f_t/\delta t.
\end{equation}
The linear fitting to data in Fig.~\ref{ss0355-evol} gives the arclet
A moving across the conjugate time axis at a rate of 0.0027(9) per
hour, and arclet B at a rate of 0.0034(13) per hour, again here the
numbers in brackets are uncertainties for the last digit. Both the
rates agree with each other and correspond to the angular velocities
of $\delta \theta_x/\delta t \simeq$ 14(5) and 18(7) mas yr$^{-1}$.
They are consistent with pulsar proper motions obtained by VLBI
measurement of (9.2, 8.2) mas yr$^{-1}$ in the right ascension and
declination \citep{ccv+04}. The change of the angular position over
the entire 12 hours of observations implies that the dense medium
structure responsible for the scattering is much larger than 0.02~AU.

\section{Conclusions}

We have carried out long observation sessions to observe the
scintillations of ten pulsars at S-band by using the Jiamusi 66-m
telescope. The newly observed dynamic spectra were mostly at the
highest frequencies for these pulsars, and some of them are the first
dynamic spectra ever published. The decorrelation bandwidths and time
scales of diffractive scintillation are derived from fitting to the
main peak of autocorrelation functions of dynamic
spectra. Well-defined parabolic arcs have been detected in the
secondary spectra of some sessions of PSRs B0355+54, B0540+23 and
B2154+40, which were used to determine the locations of the scattering
screens. The evolution of inverted arclets in the secondary spectrum
of PSR B0355+54 was observed, the angular velocity estimated from
which was consistent with VLBI measurement.

Our measurements show that scintillation parameters vary from session
to session. The frequency dependencies of both $\Delta t_d$ and
$\Delta \nu_d$ imply that the turbulence feature of the interstellar
medium deviates from the Kolmogorov turbulence. It is natural that the
intervening medium cannot be so ideally turbulent. However, the thin
screen model still holds well for PSR B1933+16. The scintillation
velocities are only a rough indication of the pulsar velocities.

Data for dynamic spectra of all pulsars presented in this paper
are available at http://zmtt.bao.ac.cn/psr-jms/.

\begin{acknowledgements}
%
%
We thank the staff members from Jiamusi deep space station and group
members in NAOC for carrying out so many long observations. The
authors thank Prof. W.~A. Coles and the referee, Prof. Dan Stinebring,
for careful reading and helpful comments. This work is partially
supported by the National Natural Science Foundation of China (Grant
No. 11403043, 11473034), the Young Researcher Grant of National
Astronomical Observatories Chinese Academy of Sciences, the Key
Research Program of the Chinese Academy of Sciences (Grant
No. QYZDJ-SSW-SLH021), the strategic Priority Research Program of
Chinese Academy of Sciences (Grant No. XDB23010200), the Open Fund of
the State Key Laboratory of Astronautic Dynamics of China (Grant
No. 2016ADL-DW0401) and the Open Project Program of the Key Laboratory
of FAST, NAOC, Chinese Academy of Sciences.

\end{acknowledgements}

\bibliographystyle{aa}
\bibliography{psr_DS}


\end{document}